\RequirePackage{lineno}
\documentclass[aps,twocolumn,showpacs,byrevtex,reprint,prd,nofootinbib]{revtex4-1}

\usepackage{graphicx}
\usepackage{dcolumn}
\usepackage{bm}
\usepackage{rotating}
\usepackage{epstopdf}
\usepackage{color}
\usepackage{verbatim}
\usepackage{multirow}
\usepackage[abs]{overpic}
\usepackage{amsmath}
\usepackage{amssymb}
\usepackage{xspace}
\usepackage[titletoc]{appendix}
\usepackage{float}
\usepackage[colorlinks,
            linkcolor=blue,
            anchorcolor=blue,
            citecolor=blue]{hyperref}

\begin{document}
\graphicspath{{figure/}}
\DeclareGraphicsExtensions{.eps,.png,.ps}
\title{\boldmath Study of dynamical dark energy using DESI DR2 BAO, CMB, and Type Ia Supernovae data}

\author{Pin Wang}
\author{Ping Wang}
\author{Jielei Zhang}
\email{Corresponding author: zhangjielei@henu.edu.cn}
\affiliation{School of Physics and Electronics, Henan University, Kaifeng 475004, China}

\begin{abstract}
Dark Energy Spectroscopic Instrument (DESI) Collaboration reported an evidence for dynamical dark energy model using baryon acoustic oscillation (BAO) measurements from DESI Data Release 2 (DR2), cosmic microwave background (CMB), and three Type Ia Supernovae (SNe) data (Pantheon+, Union3, and DES-SN5YR) recently. The significances are from $2.8\sim4.2\sigma$ depending on which SNe sample is used, and the highest significance is obtained based on the DES-SN5YR data. But more recently, Dark Energy Survey (DES) Collaboration recalibrated the DES-SN5YR data, and a significance of only $3.2\sigma$ is obtained based on the calibrated data, which is called DES-Dovekie. In this work, we perform a comprehensive analysis of five representative dynamical dark energy parametrization models, including Chevallier-Polarski-Linder (CPL), Barboza-Alcaniz (BA), exponential (EXP), logarithmic (LOG), and Jassal-Bagla-Padmanabhan (JBP), using DESI DR2 BAO, \textit{Planck} CMB, ACT DR6, and SNe data from Pantheon+, Union3, and DES-Dovekie. Similarly, a significance of around $3\sigma$ is obtained based on the DES-Dovekie data. And the significances are $2.1\sim3.7\sigma$ for different SNe samples and different parametrization models, indicating an evolutionary dark energy dynamics. Furthermore, a phantom-like behavior in the past and has transitioned into quintessence-like behavior today for the dynamical dark energy is preferred.
\end{abstract}

\maketitle

\section{Introduction}
In 1998, the accelerated expansion of the universe is observed firstly through Type Ia Supernovae~\cite{1998a, 1998b}, which is a major revolution in cosmology. This phenomenon indicates that a component with negative pressure must exist, and the component is called as ``dark energy" presently~\cite{DEa, DEb}. To explain cosmological measurements in a unified framework, a standard cosmological model ($\Lambda$CDM) is proposed: a spatially flat universe with present-time energy density composed of about 5\% baryonic matter, 25\% cold dark matter (CDM), 70\% dark energy in the form of Einstein's cosmological constant ($\Lambda$) and smaller contributions provided by massive neutrinos and radiation. Meanwhile, this model is further supported and confirmed by increasingly accurate cosmic microwave background radiation data from WMAP~\cite{wmap1, wmap2} and \textit{Planck}~\cite{planck1, planck2}. The $\Lambda$CDM is very concise, and can describe more and more observed cosmological measurements, including Type Ia Supernovae~\cite{Pantheon+1, Pantheon+2, union3, desy5}, cosmic microwave background~\cite{act, spt}, baryon acoustic oscillation~\cite{SDSS1, SDSS2, BOSS, eBOSS}, and large-scale structure~\cite{LS1, LS2, LS3, LS4}. Therefore, the $\Lambda$CDM is gradually being widely accepted.

The existence of dark energy is indisputable, but its nature remains one of the most profound unsolved problems in modern cosmology and fundamental physics. Although the $\Lambda$CDM has been successful in explaining observational data, it still faces some issues, including the ``fine-tuning" and ``cosmic coincidence" problems~\cite{problem1, problem2, problem3, problem4, problem5, problem6}. Furthermore, with the improvement of measurement accuracy, some tensions in measurements have appeared within the $\Lambda$CDM, such as the ``Hubble tension"~\cite{H01, H02, H03, H04} and the ``$S_8$ tension"~\cite{LS2, S81, S82, S83}. The $\Lambda$CDM has been challenged by these tensions, and may indicate new physics beyond $\Lambda$CDM~\cite{review1}. This leads to the necessity of exploring some alternative new cosmological models.

Especially in 2024, DESI Collaboration reported an evidence for dynamical dark energy model based on the new baryon acoustic oscillation measurements from the first data release~\cite{DESI1}. The evidence is further enhanced with the recent DR2 data from DESI~\cite{DESI2, DESI3}, which is from $2.8\sim4.2\sigma$ depending on the specific Type Ia Supernovae sample used, and the highest significance $4.2\sigma$ is obtained based on the DES-SN5YR data. Therefore, the $\Lambda$CDM has been more severely challenged and the dynamical dark energy model is preferred by these new accurate measurements. Interestingly, the results from DESI suggest a transition in the equation-of-state parameter of dark energy, which is $w<-1$ at high redshift and $w>-1$ at low redshift. Such a ``phantom-crossing" behavior is consistent with quintom dark energy model~\cite{quintom1, quintom2, quintom3}, but further confirmation is still needed. These findings have sparked extensive debates on the nature of dark energy, including phenomenological dark energy models~\cite{model1, model2, model3, model4, model5, model6, model7, model8, model9, model10, model11}, scalar field dark energy models~\cite{field1, field2, field3, field4, field5, field6, field7, field8, field9, field10}, interaction dark energy models~\cite{interaction1, interaction2, interaction3, interaction4, interaction5, interaction6, interaction7, interaction8, interaction9, interaction10, interaction11, interaction12, interaction13, interaction14, interaction15, interaction16, interaction17, interaction18, interaction19, interaction20}, early dark energy models~\cite{early1, early2, early3, early4, early5, early6, early7, early8}, holographic dark energy~\cite{holo1, holo2, holo3, holo4, holo5, holo6, holo7, holo8, holo9, holo10, holo11, holo12, holo13}, and modified gravity models~\cite{gr1, gr2, gr3, gr4, gr5, gr6, gr7, gr8, gr9, gr10}.

Undoubtedly, once the indication of dynamic dark energy reported by DESI is confirmed, it will be a major breakthrough in cosmology, with significant implications for the understanding of our universe. Therefore, given the importance of this result, numerous studies and corss-checks are needed to further test the robustness of these findings. But recently, DES Collaboration recalibrated the DES-SN5YR data, and a significance of only $3.2\sigma$ is obtained based on the calibrated data, which is called DES-Dovekie~\cite{DES1, DES2}. According to this result, the evidence for dynamical dark energy model can not reach $4\sigma$, and is only around $3\sigma$. This weakens the significance of dynamic dark energy and increases uncertainty about it, so more works are needed to further confirm this situation. Therefore, in this work, we perform a comprehensive analysis of five representative dynamical dark energy parametrization models using the combinations of baryon acoustic oscillation data, cosmic microwave background data, and Type Ia Supernovae data. Especially, we use the new calibrated Type Ia Supernovae data from DES-Dovekie, it can give more reliable results about the evolutionary properties of dark energy.

\section{$\Lambda$CDM and dynamical dark energy models}
In a universe that is uniform and isotropic on a large scale, using Friedmann-Lema\^{\i}tre-Robertson-Walker metric, and assuming that general relativity correctly describes the dynamics of expansion, the evolution with redshift $z$ of the Hubble parameter $H(z)$ is governed by the Friedmann equation, which can be written as:
\begin{equation}
\begin{aligned}
\frac{H(z)}{H_0} = & \bigg[\Omega_{\text{bc}}(1+z)^3+\Omega_{\gamma}(1+z)^4+\Omega_{\text{K}}(1+z)^2 \\
& +\Omega_{\nu}\frac{\rho_{\nu}(z)}{\rho_{\nu,0}}+\Omega_{\text{DE}}\frac{\rho_{\text{DE}}(z)}{\rho_{\text{DE},0}} \bigg]^{1/2}.
\end{aligned}
\end{equation}
Here $\Omega_{\text{bc}}=\Omega_{\text{b}}+\Omega_{\text{c}}$, $H_0$ is the Hubble parameter today, and $\Omega_{\text{b}}$, $\Omega_{\text{c}}$, $\Omega_{\gamma}$, $\Omega_{\text{K}}$, $\Omega_{\nu}$, and $\Omega_{\text{DE}}$ are the present-time fractional energy density parameters in baryons, cold dark matter, radiation, curvature, massive neutrinos, and dark energy, respectively. All the density parameters add up to unity:
\begin{equation}
\Omega_{\text{bc}}+\Omega_{\gamma}+\Omega_{\text{K}}+\Omega_{\nu}+\Omega_{\text{DE}}=1.
\end{equation}
The present-time fractional energy density in matter is defined as $\Omega_{\text{m}}=\Omega_{\text{bc}}+\Omega_{\nu}$, which includes neutrinos because they are nonrelativistic today. The energy densities of baryons and cold dark matter scale as $(1+z)^3$, while the radiation energy density scale as $(1+z)^4$, so the scaling of the neutrino energy density transitions from $(1+z)^4$ at high redshift to $(1+z)^3$ at low redshift, when $(1+z)\sim(m_{\nu}/5\times10^{-4}~\text{eV})$~\cite{pdg}. The sum of neutrino masses determines the present-day density~\cite{neutrion}, as follows:
\begin{equation}
\Omega_{\nu}h^{2}=\frac{\sum m_{\nu}}{93.14~\text{eV}},
\end{equation}
where $h\equiv H_0/(100~\text{km}~\text{s}^{-1}~\text{Mpc}^{-1})$. When the neutrinos are relativistic at high redshift, it contributes to radiation, and the total rediation density $\Omega_{\text{R}}$ is the sum of photons and relativistic neutrinos~\cite{wmap1}:
\begin{equation}
\begin{aligned}
\Omega_{\text{R}} & =\Omega_{\gamma}\left[1+\frac{7}{8}\left(\frac{4}{11}\right)^{4/3}N_{\text{eff}}\right] \\
& =2.473\times10^{-5}h^{-2}(1+0.2271N_{\text{eff}}),
\end{aligned}
\end{equation}
where $N_{\text{eff}}$ is the effective number of neutrino species. In this work, the $\sum m_{\nu}$ and $N_{\text{eff}}$ are fixed to be $0.06~\text{eV}$ and 3.044~\cite{DESI1}. We assume a flat university, and thus set $\Omega_{\text{K}}$ to be zero, motivated by the tight constraints obtained on $\Omega_{\text{K}}$ when it is allowed to vary freely~\cite{DESI2}.

For a dark energy component with an equation-of-state parameter $w(z)=P_{\text{DE}}(z)/\rho_{\text{DE}}(z)$, where $P_{\text{DE}}(z)$ is its pressure, the energy density $\rho_{\text{DE}}(z)$ normalized to its present value evolves as:
\begin{equation}
f_{\text{DE}}(z)\equiv\frac{\rho_{\text{DE}}(z)}{\rho_{\text{DE},0}}=\text{exp}\left[3\int_0^z \left[1+w(z')\right] \frac{\text{d}z'}{1+z'} \right].
\end{equation}
The $\Lambda$CDM model assumes a cosmological constant dark energy ($w(z)=-1$) with energy density is constant in space and time, so the $f_{\text{DE}}(z)$ is unity. While for any constant value $w$ of $w(z)$, the $f_{\text{DE}}(z)$ becomes $(1+z)^{3(1+w)}$, which is called $w$CDM model.

In this work, we perform a comprehensive analysis of five representative dynamical dark energy parametrization models, which are described as follows:

(1) Chevallier-Polarski-Linder (CPL) parametrization is a commonly used parametric model expresses $w(z)$ in terms of the expansion factor $a=(1+z)^{-1}$~\cite{CPL1, CPL2}, which is:
\begin{equation}
w(z)=w_0+w_a(1-a)=w_0+w_a\frac{z}{1+z},
\end{equation}
so that $w(z)$ evolves from a value $\sim(w_0+w_a)$ at high redshift to a present-day value of $\sim w_0$. This parametrization accurately represents the behavior of many physically motivated dark energy models, though more complicated evolution is possible. In the CPL model, the $f_{\text{DE}}(z)$ can be obtained as:
\begin{equation}
f_{\text{DE}}(z)=(1+z)^{3(1+w_0+w_a)}e^{-3w_a\frac{z}{1+z}}.
\end{equation}

(2) Barboza-Alcaniz (BA) parametrization is expressed as~\cite{BA1}:
\begin{equation}
w(z)=w_0+w_a\frac{1-a}{a^2+(1-a)^2}=w_0+w_a\frac{z(1+z)}{1+z^2},
\end{equation}
this parameterization maintains a similar behavior with CPL parameterization at both high and low redshifts, and showing linear behavior at low redshift while allowing deviations from CPL model. The $f_{\text{DE}}(z)$ in the BA parametrization can be calculated as:
\begin{equation}
f_{\text{DE}}(z)=(1+z)^{3(1+w_0)}(1+z^2)^{\frac{3}{2}w_a}.
\end{equation}

(3) Exponential (EXP) parametrization is given by~\cite{EXP1, EXP2}:
\begin{equation}
w(z)=w_0-w_a+w_{a}e^{1-a}=w_0+w_{a}(e^{\frac{z}{1+z}}-1),
\end{equation}
this parameterization includes an exponential term, which shows a similar behavior with CPL at low redshift according to Taylor expansion, while tends towards a value $\sim(w_0+w_a(e-1))$ at high redshift. The $f_{\text{DE}}(z)$ in the EXP parametrization can be expressed as:
\begin{equation}
f_{\text{DE}}(z)=(1+z)^{3(1+w_0-w_a)}e^{3w_aF(z)},
\end{equation}
where $F(z)=\int_{0}^{z} e^{\frac{z'}{1+z'}}\frac{\text{d}z'}{1+z'}$, and $F(z)$ can not be expressed analytically and can only be calculated numerically.

(4) Logarithmic (LOG) parametrization is expressed as~\cite{LOG1}:
\begin{equation}
w(z)=w_0-w_{a}\text{ln}(a)=w_0+w_{a}\text{ln}(1+z),
\end{equation}
this parameterization includes a logarithmic term, it also shows a similar behavior with CPL at low redshift according to Taylor expansion, while the amplitude of $w(z)$ can be very large at high redshift. The $f_{\text{DE}}(z)$ in the LOG parametrization can be calculated as:
\begin{equation}
f_{\text{DE}}(z)=(1+z)^{3(1+w_0)}e^{\frac{3}{2}w_a\left[\text{ln}(1+z)\right]^2}.
\end{equation}

(5) Jassal-Bagla-Padmanabhan (JBP) parametrization is described as~\cite{JBP1}:
\begin{equation}
w(z)=w_0+w_{a}a(1-a)=w_0+w_a\frac{z}{(1+z)^2},
\end{equation}
this parameterization includes both linear and quadratic terms in the evolution term, which shows a similar behavior with CPL at low redshift, while a difference in the dark energy behavior at high redshift. Interestingly, there are two zero points in the evolution term, so $w(z)$ tends towards a same value $\sim w_0$ at both high and low redshifts. In the JBP model, the $f_{\text{DE}}(z)$ can be obtained as:
\begin{equation}
f_{\text{DE}}(z)=(1+z)^{3(1+w_0)}e^{\frac{3}{2}w_a\frac{z^2}{(1+z)^2}}.
\end{equation}

\section{Datasets and methodology}

\subsection{Baryon acoustic oscillation}
Baryon acoustic oscillation (BAO) provides a powerful tool for measuring the expansion history, using a characteristic scale that is imprinted on matter clustering by pressure waves that propagate in the coupled baryon-photon fluid of the prerecombination Universe. DESI Collaboration has released the most accurate BAO measurements from more than 14 million galaxies and quasars based on three years of operation~\cite{DESI2}, which are extracted using various tracers including the bright galaxy survey (BGS), luminous red galaxies (LRG1-3), emission line galaxies (ELG1-2), quasars (QSO), and Lyman-$\alpha$ (Ly$\alpha$) forest quasars. These measurements give the values of $D_{\text{M}}(z)/r_{\text{d}}$, $D_{\text{H}}(z)/r_{\text{d}}$, and $D_{\text{V}}(z)/r_{\text{d}}$ at effective redshifts $z_{\text{eff}}$ ranging from 0.295 to 2.330, where $D_{\text{M}}(z)=c\int_0^z \frac{\text{d}z'}{H(z')}$ is the transverse comoving distance, $D_{\text{H}}(z)=\frac{c}{H(z)}$ is the line-of-sight direction distance, $D_{\text{V}}(z)=\left[zD_{\text{M}}(z)^2D_{\text{H}}(z)\right]^{\frac{1}{3}}$ is the angle-average distance, and $r_{\text{d}}$ is the sound horizon at the drag epoch. So in this work, we use these BAO measurements from DESI Data Release 2 (DR2)\footnote{\url{https://github.com/CobayaSampler/bao_data}}.

\subsection{Cosmic microwave background}
Since the discovery of cosmic microwave background (CMB), its temperature ($TT$) anisotropy has become one of the most powerful ways of studying cosmology and the physics of the early Universe ($z_*\approx1090$). With the improvement of measurement accuracy, the polarization ($EE$) and cross ($TE$) power spectra are also measured. Just as BAO, the same physics imprints the acoustic peaks in the CMB power spectra, and the angular scale of these peaks is measured with exquisite precision. CMB can provide tight constraints to $r_{\text{d}}$, so the BAO+CMB combination allows absolute measurements of $D_{\text{M}}(z)$ and $D_{\text{H}}(z)$.
Now the power spectra of anisotropies in the CMB have been exquisitely measured by \textit{Planck} Collaboration~\cite{planck2}. Therefore, we make full use of the $TT$, $EE$, and $TE$ power spectra from \textit{Planck}\footnote{\url{https://github.com/CobayaSampler/planck_native_data}}, specifically using the \texttt{simall}, \texttt{Commander} (for $\ell<30$), and \texttt{CamSpec} (for $\ell\geq30$) likelihoods, plus the combination of \textit{Planck} and ACT DR6 CMB lensing\footnote{\url{https://lambda.gsfc.nasa.gov/product/act/actadv_dr6_lensing_lh_get.html}} from Ref.~\cite{act}.

\subsection{Type Ia supernova}
Type Ia supernovae (SNe) are luminous standardizable candles that provide another probe of the expansion history of the Universe, especially useful at low redshifts ($0.01<z<0.3$) where BAO measurements are limited by cosmic variance. Three SNe\footnote{\url{https://github.com/CobayaSampler/sn_data}} datasets are used in this work: Pantheon+~\cite{Pantheon+1, Pantheon+2}, Union3~\cite{union3}, and DES-Dovekie~\cite{desy5}. The Pantheon+ sample comprises 1701 light curves of 1550 spectroscopically confirmed SNe in the redshift range $0.001<z<2.26$. The Union3 compilation has 2087 SNe in the redshift range $0.01<z<2.26$, while 1363 of which are common to Pantheon+, though the analysis methodologies are substantially different. Finally, the DES-Dovekie dataset consists of 1623 photometrically identified SNe from the full DES five-year supernova survey with redshifts in the range $0.1<z<1.13$, together with a small set of 197 SNe at $0.025<z<0.1$ from historical low redshift samples, some of which are in common with the Pantheon+ and Union3 datasets. It is important to note that in all SNe samples, the absolute magnitude $M$ of the SNe is completely degenerate with the Hubble parameter $H_0$, so the distance moduli can be scaled by any constant offset. Therefore, for cosmological inference, we marginalize over the value of $M$.

\subsection{Methodology}
In this work, we utilize the Markov Chain Monte Carlo (MCMC) sampling to explore the parameter space using the Metropolis-Hastings algorithm~\cite{mcmc1, mcmc2} as implemented in \texttt{Cobaya}~\cite{cobaya}. Theory models are computed using interfaces to the Boltzmann solver CAMB~\cite{camb1, camb2} and \texttt{Cobaya} likelihoods. The convergence criterion for MCMC sampling is that the Gelman-Rubin statistic~\cite{RB} satisfies $R-1<0.02$. Summary statistics for our chains as well as plots are obtained with \texttt{getdist} software package~\cite{getdist}. In order to determine the best fit points and the corresponding $\chi^2$, we use the \texttt{BOBYQA}~\cite{BOYA1, BOYA2, BOYA3} algorithm starting from the maximum a posteriori (MAP) points of each of the chains in the MCMC sampling. When comparing the fits of two different models, we compare the quantity $\Delta\chi^2_{\text{MAP}}=-2\Delta\text{ln}\mathcal{L}$ representing twice the difference in the negative log posteriors at the maximum posterior points for each model. We also employ the deviance information criterion (DIC)~\cite{DIC1, DIC2} to perform statistical comparisons among the different models. The DIC is defined as
\begin{equation}
    {\rm DIC} \equiv D(\bar{\theta})+2p_D,
    \quad
    p_D \equiv \left\langle D(\theta)\right\rangle-D(\bar{\theta}),
\end{equation}
where $\theta$ denotes the set of model parameters, $\bar{\theta}$ denotes the posterior mean parameter point, $\langle D(\theta)\rangle$ is the mean deviance evaluated over the MCMC chains, and $p_D$ denotes the effective number of parameters, which measures the effective complexity of the model. Here, $D(\theta)=-2\ln\mathcal{L}(\theta)+C$ is the deviance, where $C$ is a constant independent of the model parameters. The difference between each model and the $\Lambda$CDM model is defined as
\begin{equation}
    \Delta{\rm DIC}_i \equiv {\rm DIC}_i-{\rm DIC}_{\Lambda{\rm CDM}}.
\end{equation}
Therefore, a smaller DIC value indicates that the corresponding model is more strongly supported by the data. The free cosmological parameters of these models and the uniform priors used are listed in Table~\ref{tab:prior}. The parameters for the $\Lambda$CDM model are $\theta_{\Lambda\text{CDM}}=\left\{\Omega_{b}h^2, \Omega_{c}h^2, 100\theta_{\text{MC}}, \text{ln}(10^{10}A_s), n_s, \tau\right\}$, while the parameters for $w\Lambda$CDM model and dynamical dark energy models are $\theta_{w\Lambda\text{CDM}}=\left\{\theta_{\Lambda\text{CDM}}, w\right\}$ and $\theta=\left\{\theta_{\Lambda\text{CDM}}, w_0, w_a\right\}$.

\begin{table}[htbp]
\renewcommand{\arraystretch}{1.25}
\begin{center}
\caption{Cosmological parameters and priors used for different models in this work, and all the priors are flat in the given ranges.}
\label{tab:prior}
\begin{tabular}{lcc}
  \hline
  \hline
  Model \qquad \qquad \qquad \qquad \qquad & \ \ \ \ Parameter \ \ \ \ & \ \ \ \ Prior \ \ \ \ \\
  \hline
  \textbf{$\boldsymbol{\Lambda}$CDM}  & $\Omega_{b}h^2$ & $\mathcal{U}[0.005, 0.1]$  \\
                                      & $\Omega_{c}h^2$ & $\mathcal{U}[0.001, 0.99]$ \\
                                      & $100\theta_{\text{MC}}$ & $\mathcal{U}[0.5, 10]$ \\
                                      & $\text{ln}(10^{10}A_s)$ & $\mathcal{U}[1.61, 3.91]$ \\
                                      & $n_s$  & $\mathcal{U}[0.8, 1.2]$ \\
                                      & $\tau$ & $\mathcal{U}[0.01, 0.8]$ \\
  \textbf{$\boldsymbol{w\Lambda}$CDM} & $w$   & $\mathcal{U}[-3, 1]$\\
  \textbf{Dynamical dark energy}      & $w_0$ & $\mathcal{U}[-3, 1]$\\
                                      & $w_a$ & $\mathcal{U}[-6, 6]$\\
  \hline
  \hline
\end{tabular}
\end{center}
\end{table}

\section{Results and discussions}
\begin{figure*}[htbp]
\begin{center}
\includegraphics[width=0.32\textwidth]{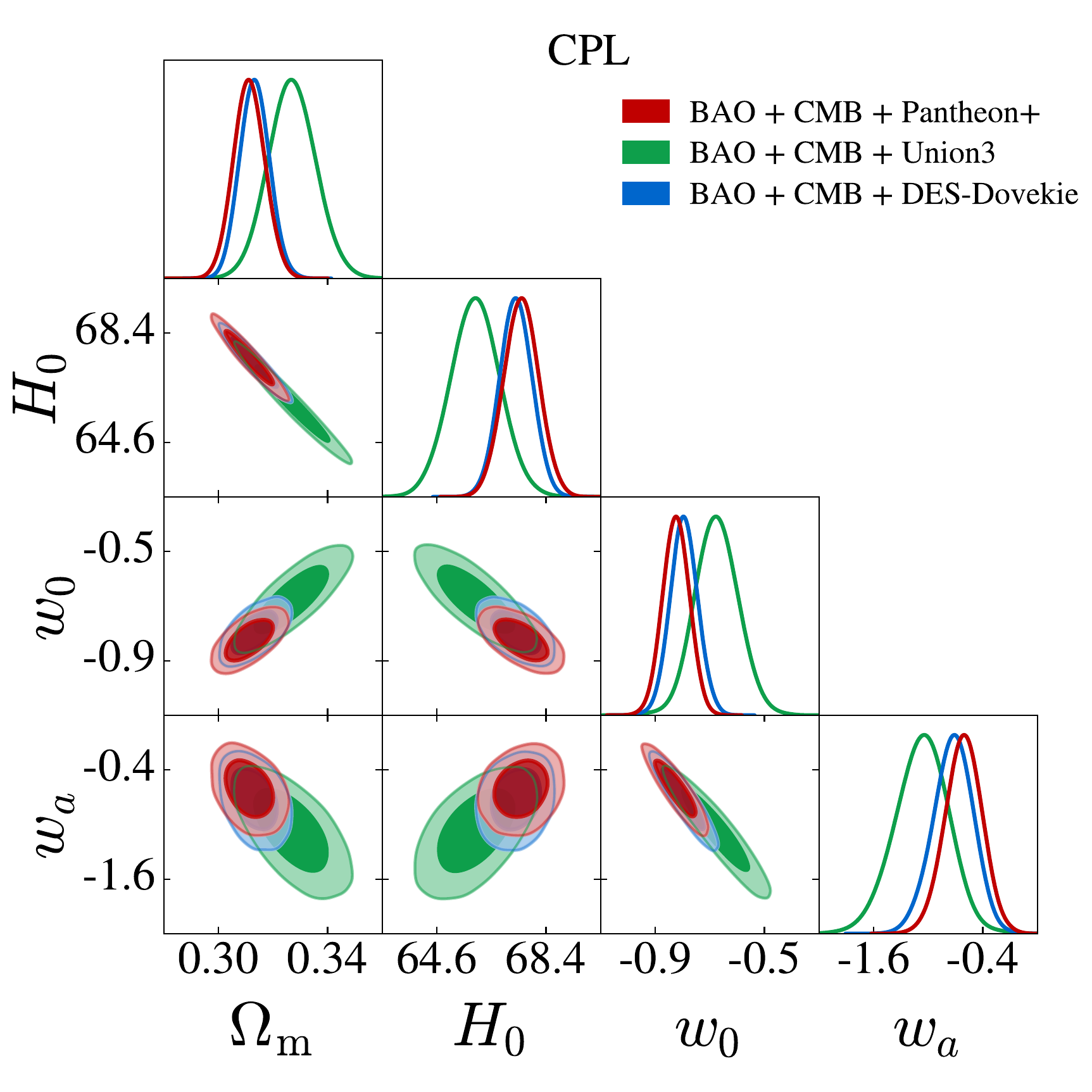}
\includegraphics[width=0.32\textwidth]{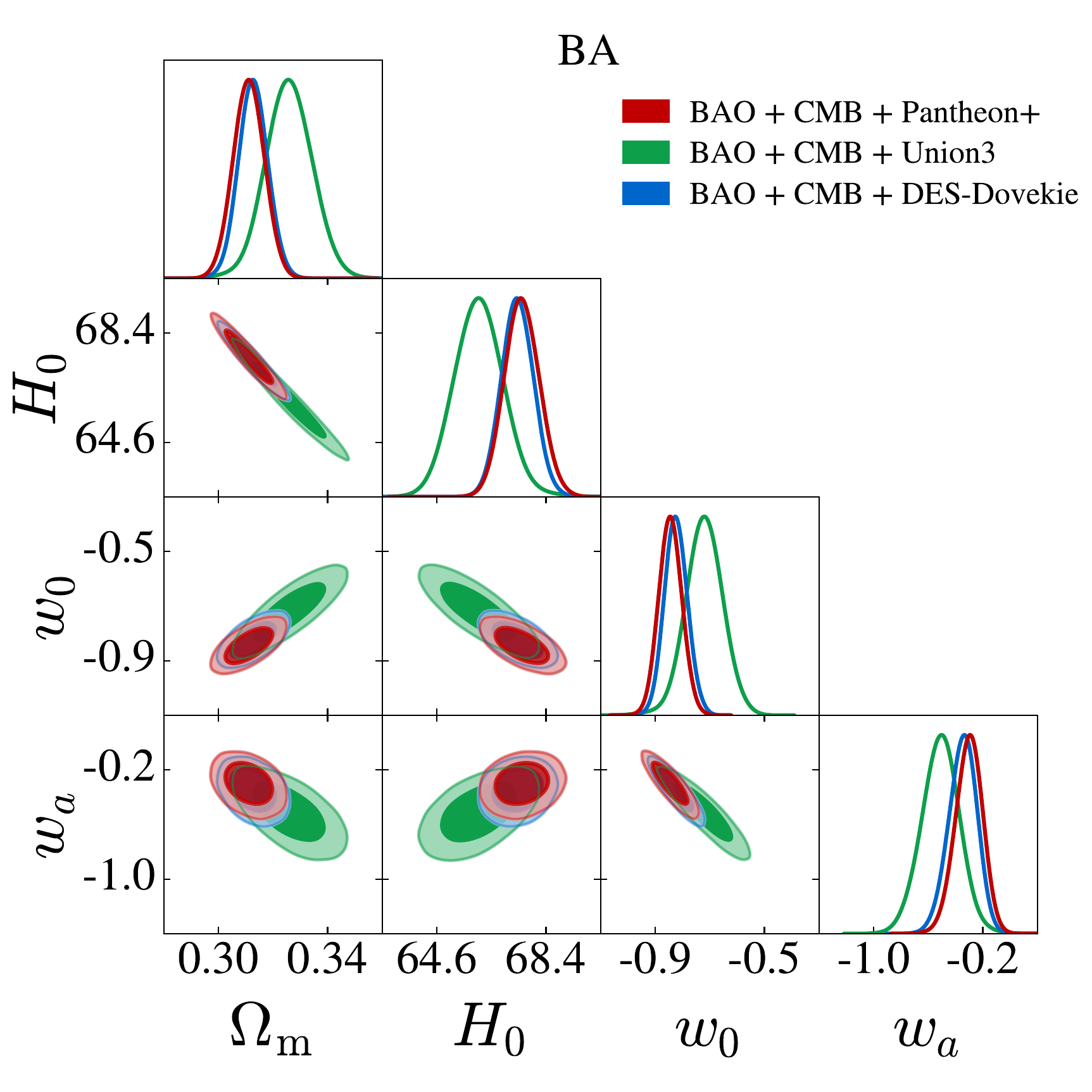}
\includegraphics[width=0.32\textwidth]{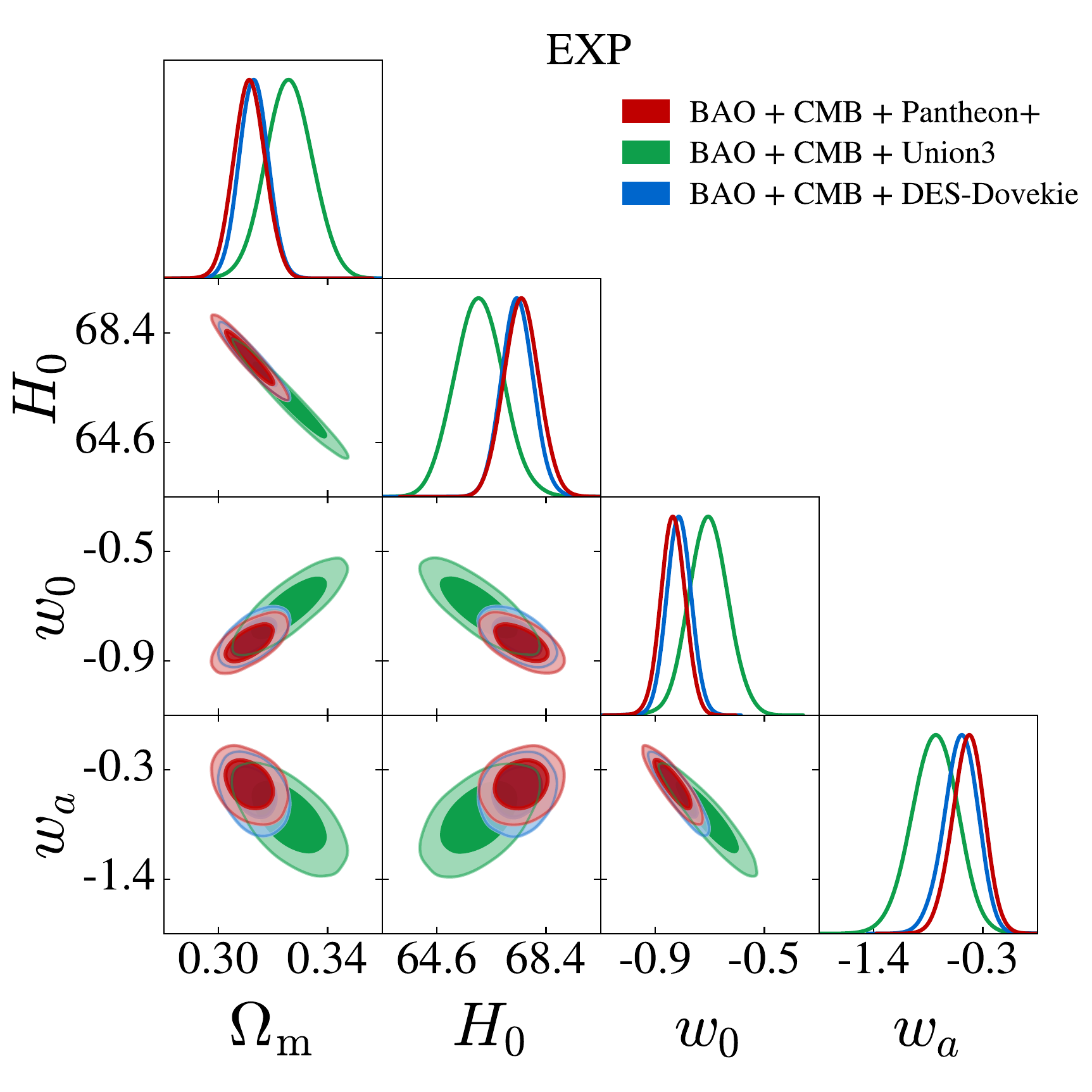}
\includegraphics[width=0.32\textwidth]{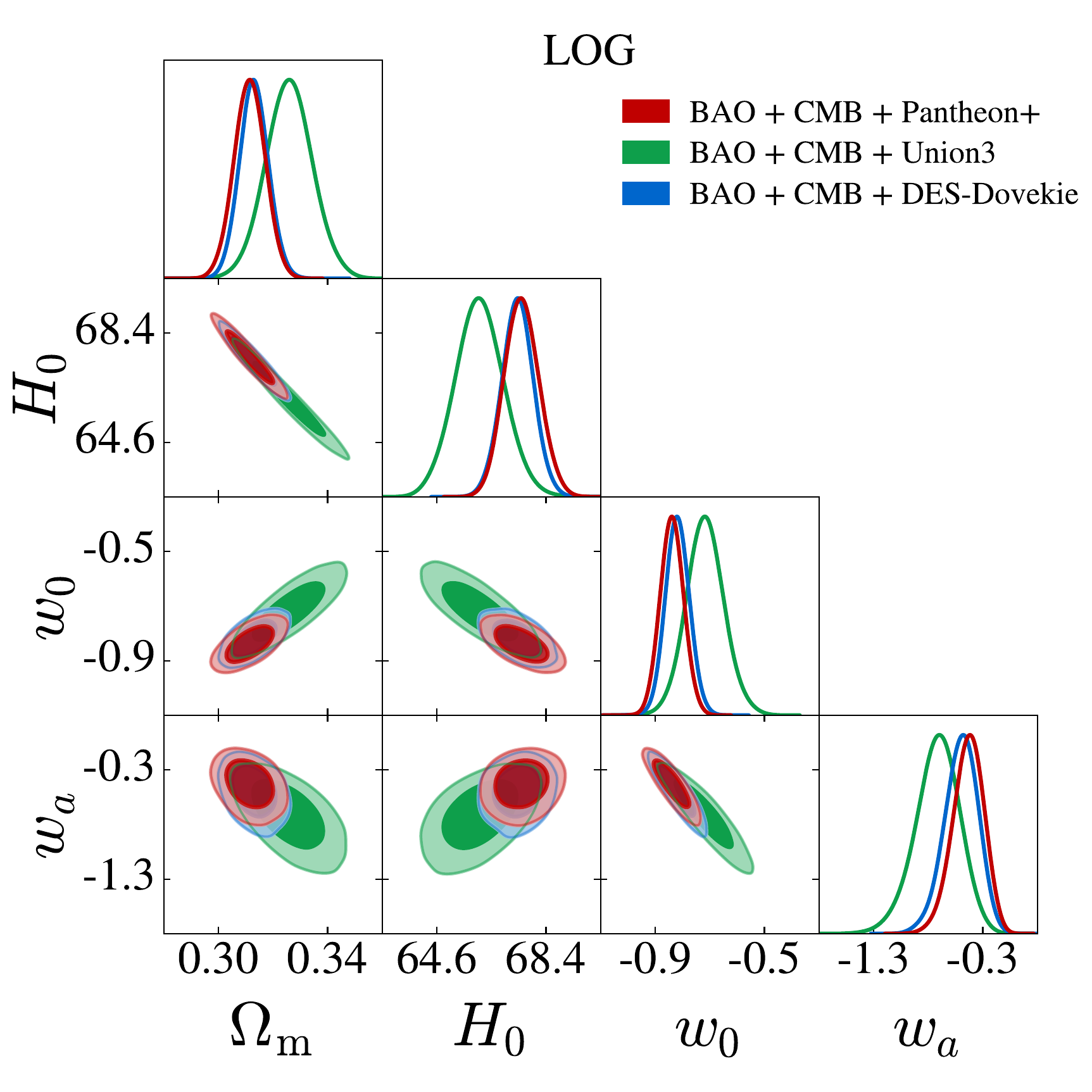}
\includegraphics[width=0.32\textwidth]{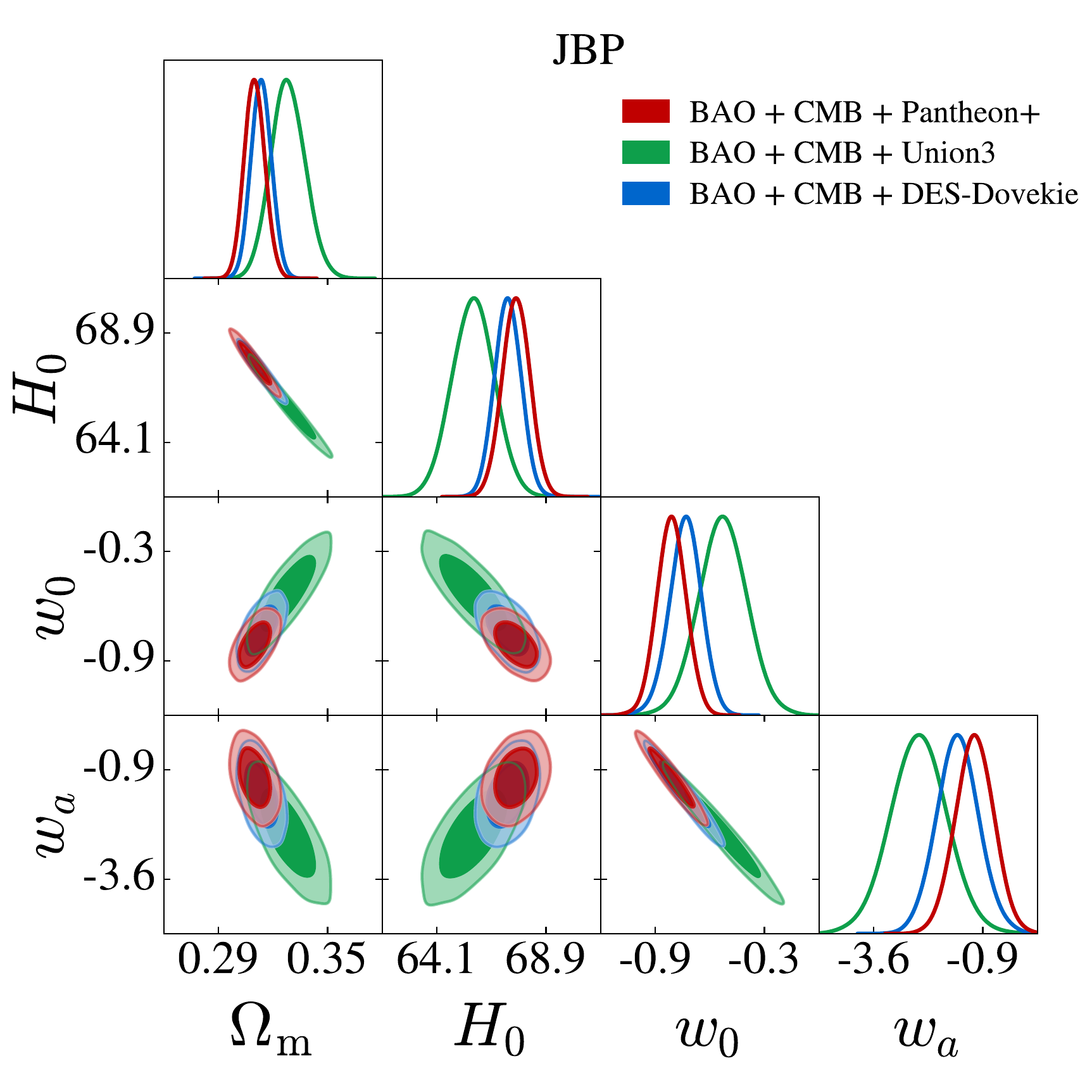}
\caption{Marginalised posteriors on $\Omega_{\text{m}}$, $H_0$, $w_0$, and $w_a$ in the CPL, BA, EXP, LOG, and JBP models, and contour plots are at the 68\% ($1\sigma$) and 95\% ($2\sigma$) credible intervals. The results are from the combinations of BAO, CMB, and different SNe from Pantheon+, Union3, and DES-Dovekie in red, green, and blue, respectively.}
\label{fig:result1}
\end{center}
\end{figure*}

In this section, we present the constraints on the cosmological parameters. Table~\ref{tab:result1} shows the marginalized constraints on cosmological parameters from different combinations of observational data, and the parameter values are the means of the one-dimensional posterior distributions. The marginalized posterior distributions of these parameters are shown in Fig.~\ref{fig:result1}.

\begin{table*}[htbp]
\renewcommand{\arraystretch}{1.25}
\begin{center}
\caption{Summary table of cosmological parameter constraints in $\Lambda$CDM, $w\Lambda$CDM, and various dynamical dark energy models. Results quoted for all parameters are the marginalized posterior means and 68\% credible intervals.}
\label{tab:result1}
\begin{tabular}{lcccc}
  \hline
  \hline
  Model/Dataset & $\Omega_{\text{m}}$ & $H_0$ (km s$^{-1}$ Mpc$^{-1}$) & $w$ or $w_0$ & $w_a$ \\
  \hline
  \textbf{$\boldsymbol{\Lambda}$CDM} \\
  BAO + CMB + Pantheon+   & $0.3037\pm0.0037$ & $68.09\pm0.28$ & ... & ... \\
  BAO + CMB + Union3      & $0.3035\pm0.0035$ & $68.11\pm0.27$ & ... & ... \\
  BAO + CMB + DES-Dovekie & $0.3040\pm0.0035$ & $68.07\pm0.27$ & ... & ... \\
  \textbf{$\boldsymbol{w}$CDM} \\
  BAO + CMB + Pantheon+   & $0.3047\pm0.0050$ & $67.96\pm0.57$ & $-0.994\pm0.023$ & ... \\
  BAO + CMB + Union3      & $0.3041\pm0.0059$ & $68.03\pm0.68$ & $-0.997\pm0.027$ & ... \\
  BAO + CMB + DES-Dovekie & $0.3061\pm0.0048$ & $67.79\pm0.53$ & $-0.988\pm0.021$ & ... \\
  \textbf{CPL} \\
  BAO + CMB + Pantheon+   & $0.3112\pm0.0057$ & $67.53\pm0.60$ & $-0.839\pm0.055$ & $-0.610\pm0.204$ \\
  BAO + CMB + Union3      & $0.3269\pm0.0087$ & $65.95\pm0.84$ & $-0.673\pm0.089$ & $-1.062\pm0.292$ \\
  BAO + CMB + DES-Dovekie \ \ \ \ & \ \ \ \ $0.3132\pm0.0054$ \ \ \ \ & \ \ \ \ $67.34\pm0.55$ \ \ \ \ & \ \ \ \ $-0.807\pm0.056$ \ \ \ \ & \ \ \ \ $-0.721\pm0.221$ \ \ \ \ \\
  \textbf{BA} \\
  BAO + CMB + Pantheon+   & $0.3111\pm0.0057$ & $67.54\pm0.60$ & $-0.863\pm0.047$ & $-0.299\pm0.099$ \\
  BAO + CMB + Union3      & $0.3258\pm0.0085$ & $66.06\pm0.84$ & $-0.722\pm0.075$ & $-0.508\pm0.136$ \\
  BAO + CMB + DES-Dovekie & $0.3127\pm0.0053$ & $67.39\pm0.55$ & $-0.840\pm0.047$ & $-0.344\pm0.104$ \\
  \textbf{EXP} \\
  BAO + CMB + Pantheon+   & $0.3113\pm0.0057$ & $67.53\pm0.60$ & $-0.852\pm0.050$ & $-0.462\pm0.152$ \\
  BAO + CMB + Union3      & $0.3259\pm0.0085$ & $66.07\pm0.83$ & $-0.706\pm0.080$ & $-0.784\pm0.220$ \\
  BAO + CMB + DES-Dovekie & $0.3129\pm0.0053$ & $67.39\pm0.54$ & $-0.827\pm0.050$ & $-0.533\pm0.163$ \\
  \textbf{LOG} \\
  BAO + CMB + Pantheon+   & $0.3115\pm0.0056$ & $67.53\pm0.60$ & $-0.856\pm0.048$ & $-0.429\pm0.143$ \\
  BAO + CMB + Union3      & $0.3259\pm0.0085$ & $66.09\pm0.83$ & $-0.720\pm0.076$ & $-0.708\pm0.200$ \\
  BAO + CMB + DES-Dovekie & $0.3130\pm0.0052$ & $67.40\pm0.53$ & $-0.834\pm0.049$ & $-0.492\pm0.154$ \\
  \textbf{JBP} \\
  BAO + CMB + Pantheon+   & $0.3098\pm0.0056$ & $67.58\pm0.60$ & $-0.810\pm0.080$ & $-1.146\pm0.474$ \\
  BAO + CMB + Union3      & $0.3281\pm0.0094$ & $65.72\pm0.92$ & $-0.529\pm0.132$ & $-2.527\pm0.694$ \\
  BAO + CMB + DES-Dovekie & $0.3135\pm0.0056$ & $67.21\pm0.58$ & $-0.733\pm0.087$ & $-1.565\pm0.519$ \\
  \hline
  \hline
\end{tabular}
\end{center}
\end{table*}

The results show that, for the same data combination, different dark energy models give similar constraints on $\Omega_m$ and $H_0$, with no significant differences within a range of two standard deviations. For the $w$CDM model, the constraints on $w$ from all three data combinations are close to $-1$, showing no significant deviation from $\Lambda$CDM. It should be noted that even if the $w$ is consistent with $-1$, it does not mean that dark energy has no dynamic behavior, but only that the weighted average of $w$ is close to $-1$ throughout the effective redshift range. From the Fig.~\ref{fig:result1}, we can also see different combinations of SNe data give the different results on the constraints of dynamic dark energy parameters, where Pantheon+ and DES-Dovekie give very similar constraints with a large overlap between their two-dimensional credible intervals, while the Union3 has slightly significant differences. This indicates that the dynamical dark energy parameters, especially $w_0$ and $w_a$, have a significant data dependence on the SNe samples used. In addition, the three sets of data show similar parametric correlations. Here, $\Omega_m$ shows a strong negative correlation with $H_0$, while $w_0$ and $w_a$ show a clear negative correlation. However, the constraints on $w_0$ and $w_a$ depend more strongly on the different models. Since $w_0$ and $w_a$ describe the evolution of the dark energy equation-of-state, we further study them by showing the two-dimensional constraints in the $w_0$-$w_a$ plane for each model.

\begin{figure*}[htbp]
\begin{center}
\includegraphics[width=0.32\textwidth]{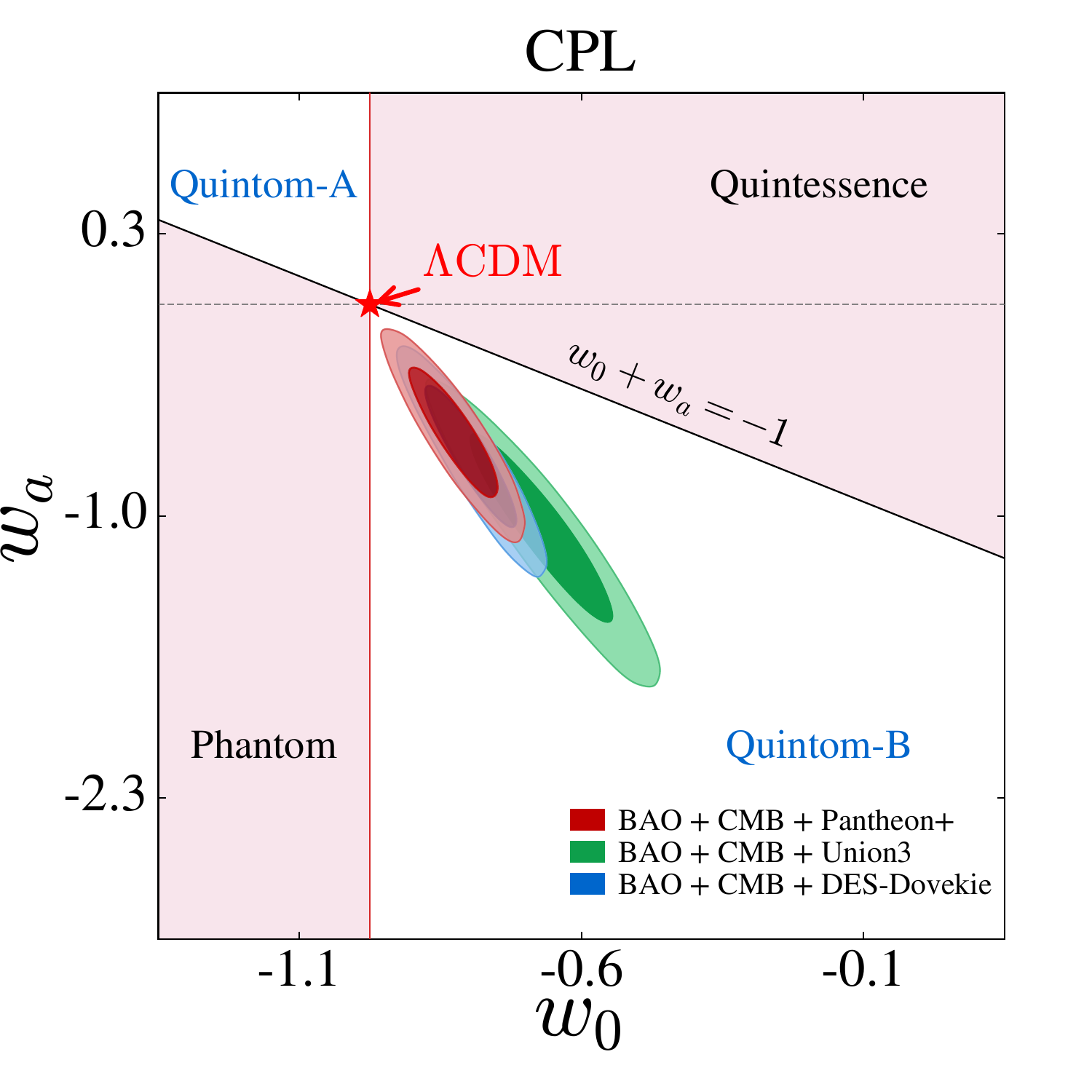}
\includegraphics[width=0.32\textwidth]{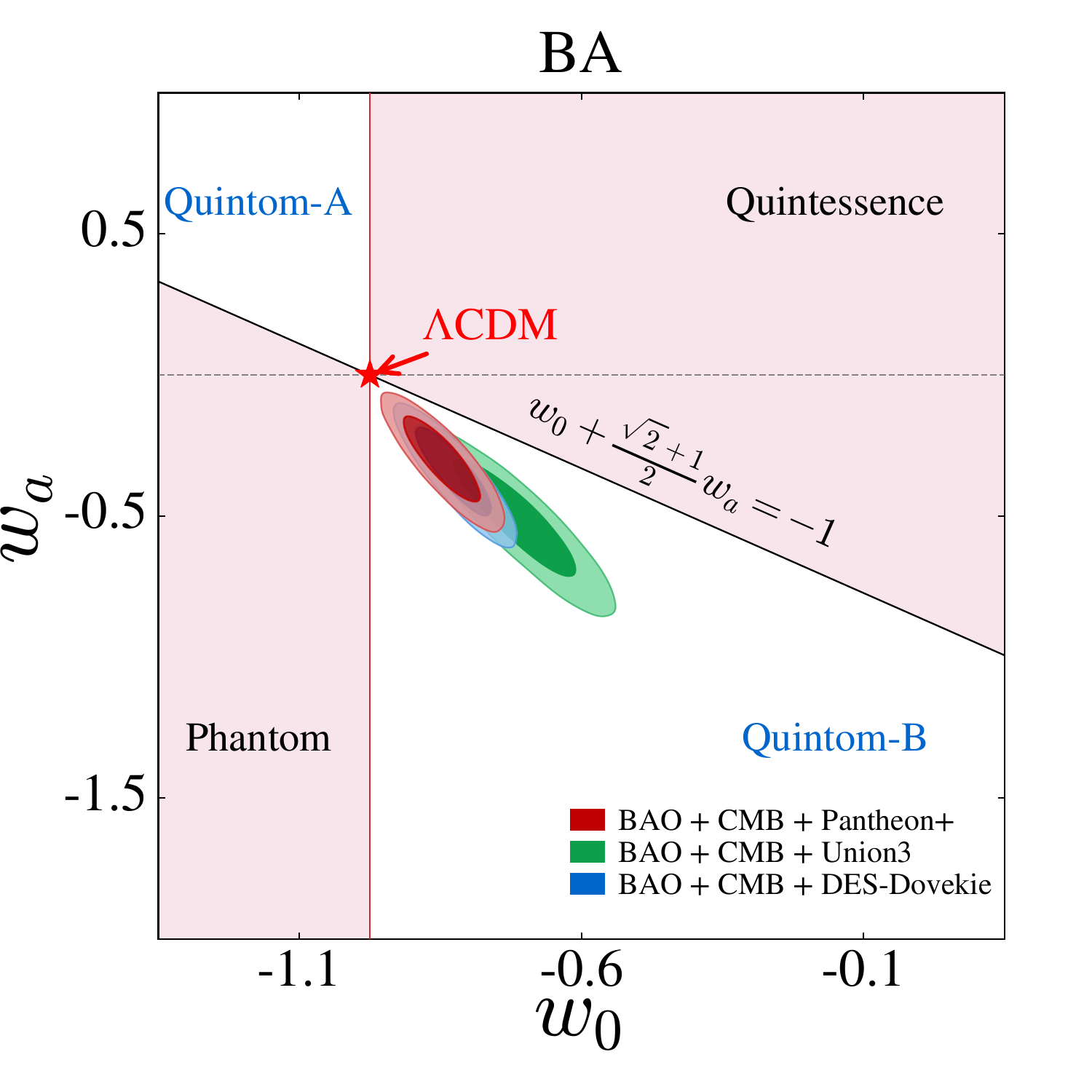}
\includegraphics[width=0.32\textwidth]{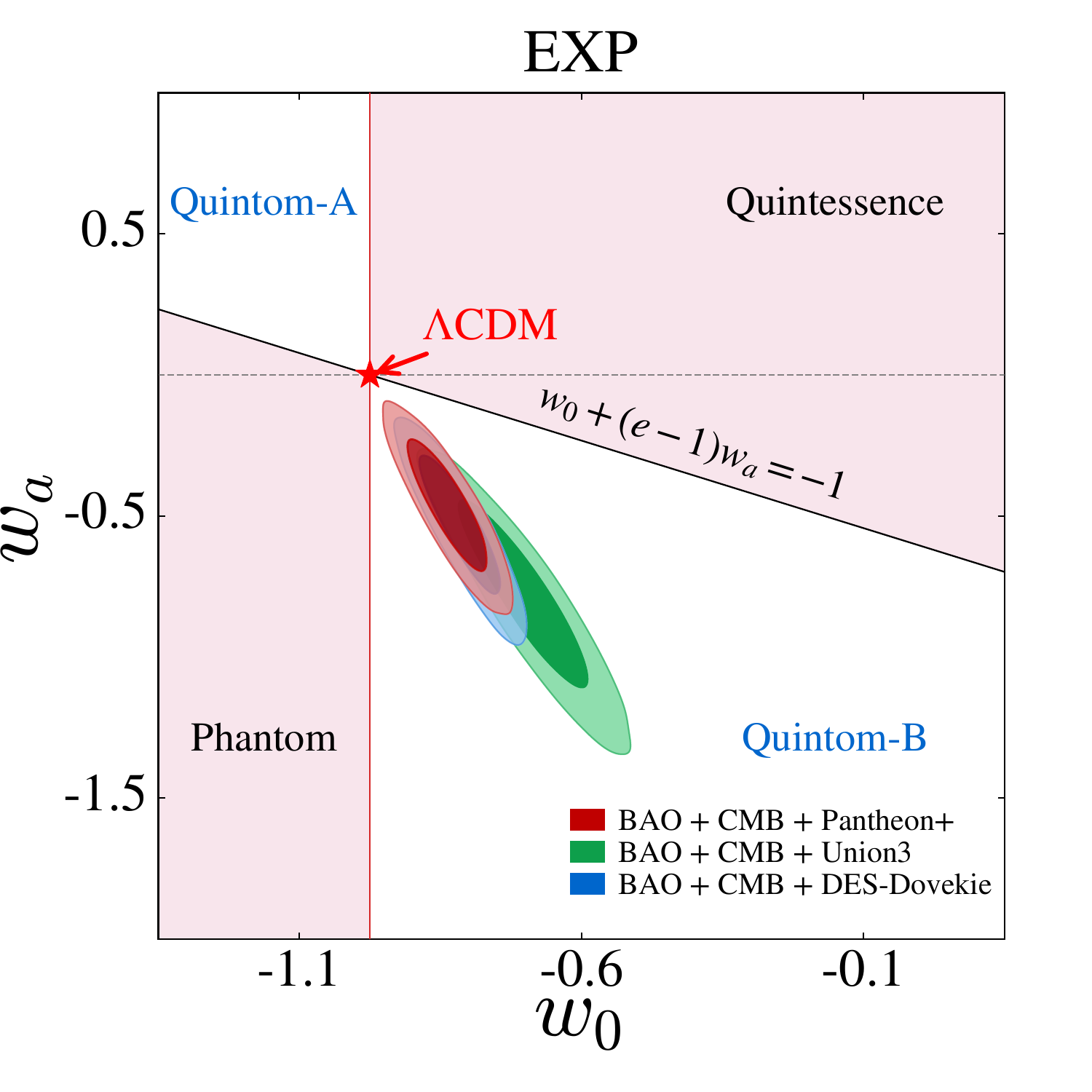}
\includegraphics[width=0.32\textwidth]{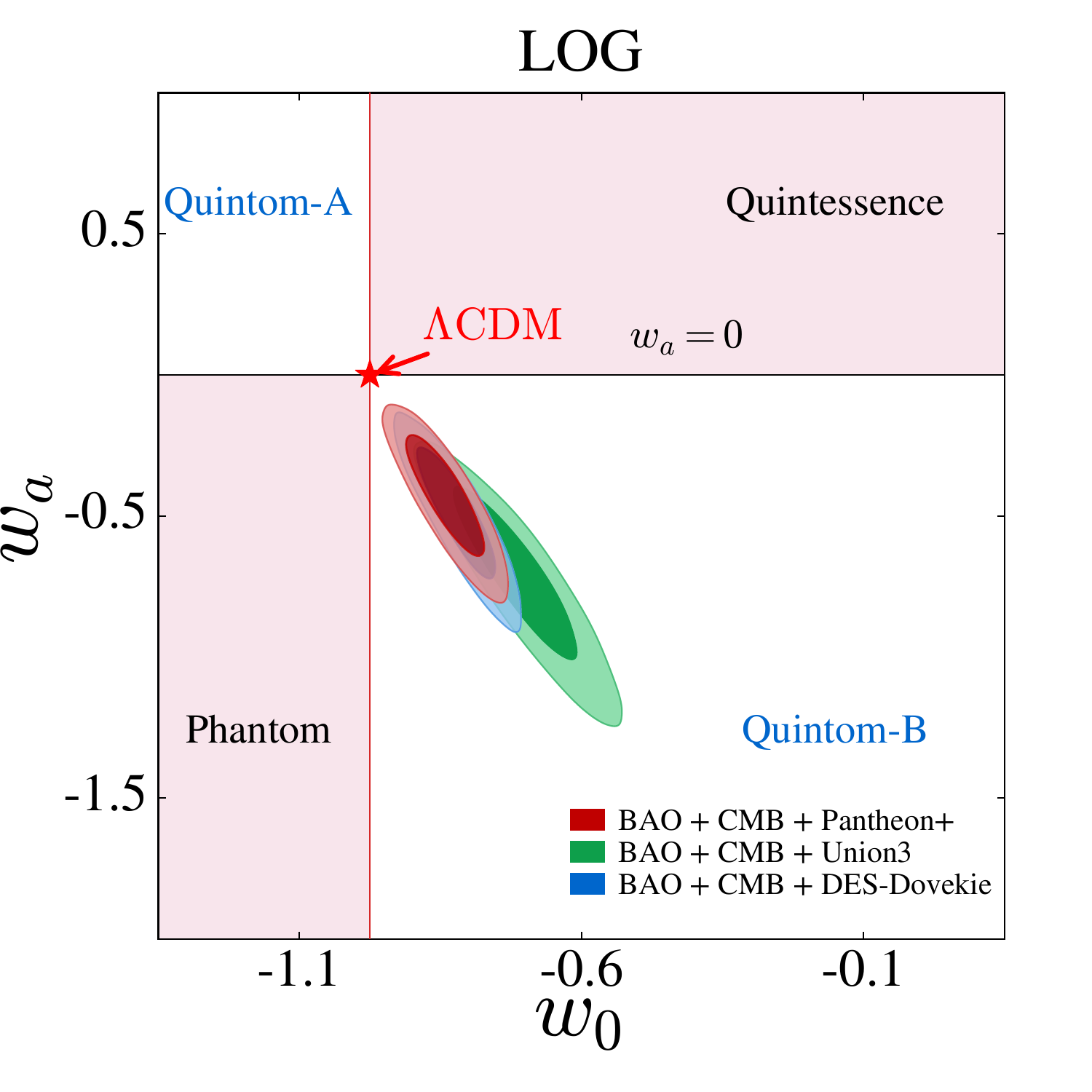}
\includegraphics[width=0.32\textwidth]{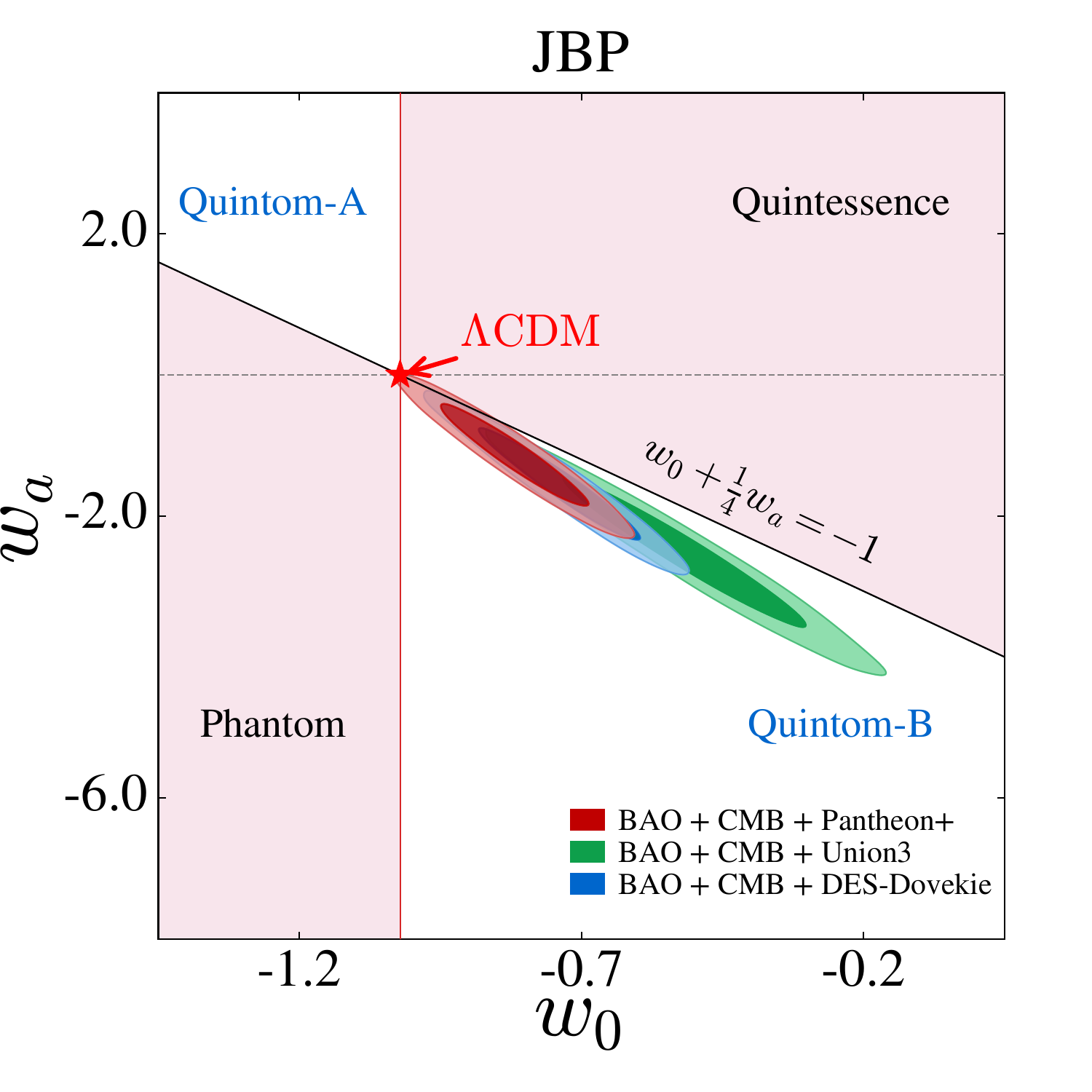}
\caption{Contour plots at the 68\% ($1\sigma$) and 95\% ($2\sigma$) credible intervals for the five dynamical dark energy models on the $w_0$-$w_a$ plane. The red pentagrams mark the $\Lambda$CDM model, and the regions of quintessence, phantom, quintom-A, and quintom-B are divided by boundary lines.}
\label{fig:w0wa}
\end{center}
\end{figure*}

Figure~\ref{fig:w0wa} shows the contour plots on the $w_0$-$w_a$ plane for the five dynamical dark energy models. For the $\Lambda$CDM model, the dark energy equation-of-state is $w(z)=-1$. Therefore, it corresponds to the point $(w_0, w_a)=(-1, 0)$ in the parameter space. Here, $w_0$ characterizes the present-day value of the dark energy equation-of-state, while $w_a$ describes the amplitude of its redshift evolution. As shown in Fig.~\ref{fig:w0wa}, $w_0$ represents the value of the dark energy equation-of-state at the present epoch, so the vertical line $w_0=-1$ divides the parameter space into the present-day quintessence-like and phantom-like regions, where $w_0>-1$ means quintessence-like behavior, while $w_0<-1$ means phantom-like behavior~\cite{Qun1}. However, the parameter $w_0$ alone can not tell us the behavior of dark energy at high redshift. For each dynamical dark energy model, we calculate its equation-of-state at high redshift and set $w(z)=-1$ to obtain another boundary line. By combining the present-day and high redshift conditions, the parameter space can be divided into four typical regions~\cite{Qun2, Qun3}. The boundary conditions and region classifications for different dynamical dark energy models are shown in Table~\ref{tab:w0wa}. The quintessence region always has $w(z)>-1$ at any time, while the phantom region always has $w(z)<-1$. The quintom-A region denotes the evolution from $w(z)>-1$ at high redshift to $w(z)<-1$ at low redshift, whereas quintom-B denotes an evolution from $w(z)<-1$ to $w(z)>-1$. In the last two cases, the dark energy equation-of-state crosses the phantom divide $w(z)=-1$ as the universe evolves.

\begin{table*}[htbp]
\renewcommand{\arraystretch}{1.25}
\begin{center}
\caption{Classification of the different dynamical dark energy models in the $w_0$-$w_a$ plane. Quintom-A denotes the dark energy evolution from $w>-1$ at high redshift to $w<-1$ at low redshift, whereas Quintom-B denotes an evolution from $w<-1$ to $w>-1$.}
\label{tab:w0wa}
\begin{tabular}{lcccc}
  \hline
  \hline
  Model & Quintessence & Phantom & Quintom-A & Quintom-B \\
  \hline
  \textbf{CPL:}                & $w_0>-1$ and & $w_0<-1$ and & $w_0<-1$ and & $w_0>-1$ and \\
  $w(z)=w_0+w_a\frac{z}{1+z}$  & $w_0+w_a>-1$ & $w_0+w_a<-1$ & $w_0+w_a>-1$ & $w_0+w_a<-1$ \\
  \\
  \textbf{BA:}                       & $w_0>-1$ and & $w_0<-1$ and & $w_0<-1$ and & $w_0>-1$ and \\
  $w(z)=w_0+w_a\frac{z(1+z)}{1+z^2}$ & $w_0+\frac{\sqrt{2}+1}{2}w_a>-1$ & $w_0+\frac{\sqrt{2}+1}{2}w_a<-1$ & $w_0+\frac{\sqrt{2}+1}{2}w_a>-1$ & $w_0+\frac{\sqrt{2}+1}{2}w_a<-1$ \\
  \\
  \textbf{EXP:}                             & $w_0>-1$ and & $w_0<-1$ and & $w_0<-1$ and & $w_0>-1$ and \\
  $w(z)=w_0+w_{a}(e^{\frac{z}{1+z}}-1)$ \ \ & \ \ $w_0+(e-1)w_a>-1$ \ \ & \ \ $w_0+(e-1)w_a<-1$ \ \ & \ \ $w_0+(e-1)w_a>-1$ \ \ & \ \ $w_0+(e-1)w_a<-1$ \ \ \\
  \\
  \textbf{LOG:}          & $w_0>-1$ and & $w_0<-1$ and & $w_0<-1$ and & $w_0>-1$ and \\
  $w(z)=w_0+w_a\ln(1+z)$ & $w_a\geq0$   & $w_a\leq0$   & $w_a>0$      & $w_a<0$ \\
  \\
  \textbf{JBP:}                   & $w_0>-1$ and & $w_0<-1$ and & $w_0<-1$ and & $w_0>-1$ and \\
  $w(z)=w_0+w_a\frac{z}{(1+z)^2}$ & $w_0+\frac{1}{4}w_a>-1$ & $w_0+\frac{1}{4}w_a<-1$ & $w_0+\frac{1}{4}w_a>-1$ & $w_0+\frac{1}{4}w_a<-1$ \\
  \hline
  \hline
\end{tabular}
\end{center}
\end{table*}

For the CPL, BA, EXP, and LOG models, their preferred parameter regions all satisfy $w_0>-1$. When the redshift increases, $w(z)$ becomes more negative, crosses the boundary $w(z)=-1$ at low or intermediate redshift, and enters the phantom-like region. Thus, these models show quintom-B behavior~\cite{Qun3}, changing from a phantom-like state in the past to a quintessence-like state at present. However, the JBP model shows a different behavior and needs to be discussed separately. Its dark energy equation-of-state is given by Eq.(14), which satisfies $w(z=0)=w_0$ and $w(z\to\infty)=w_0$. For the preferred region with $w_0>-1$ and $w_a<0$, the function $z/(1+z)^2$ has its maximum at $z=1$. Thus, $w(z)$ reaches its minimum value at this redshift:
\begin{equation}
w_{\min}=w(z=1)=w_0+\frac{1}{4}w_a.
\end{equation}
When $w_0+w_a/4<-1$, the JBP equation-of-state decreases from the present-day quintessence-like region, crosses $w(z)=-1$, and temporarily enters the phantom-like region at intermediate redshift. At higher redshift, however, $w(z)$ increases again and gradually approaches $w_0>-1$. Therefore, as the redshift increases, the JBP model evolves from a quintessence-like state to a phantom-like state, and then returns to a quintessence-like state. In conclusion, the JBP model may cross the $w(z)=-1$ boundary twice and only enter the phantom-like region temporarily. Unlike the standard quintom-B evolution, it does not remain phantom-like at the very high redshift.

\begin{figure*}[htbp]
\begin{center}
\includegraphics[width=0.32\textwidth]{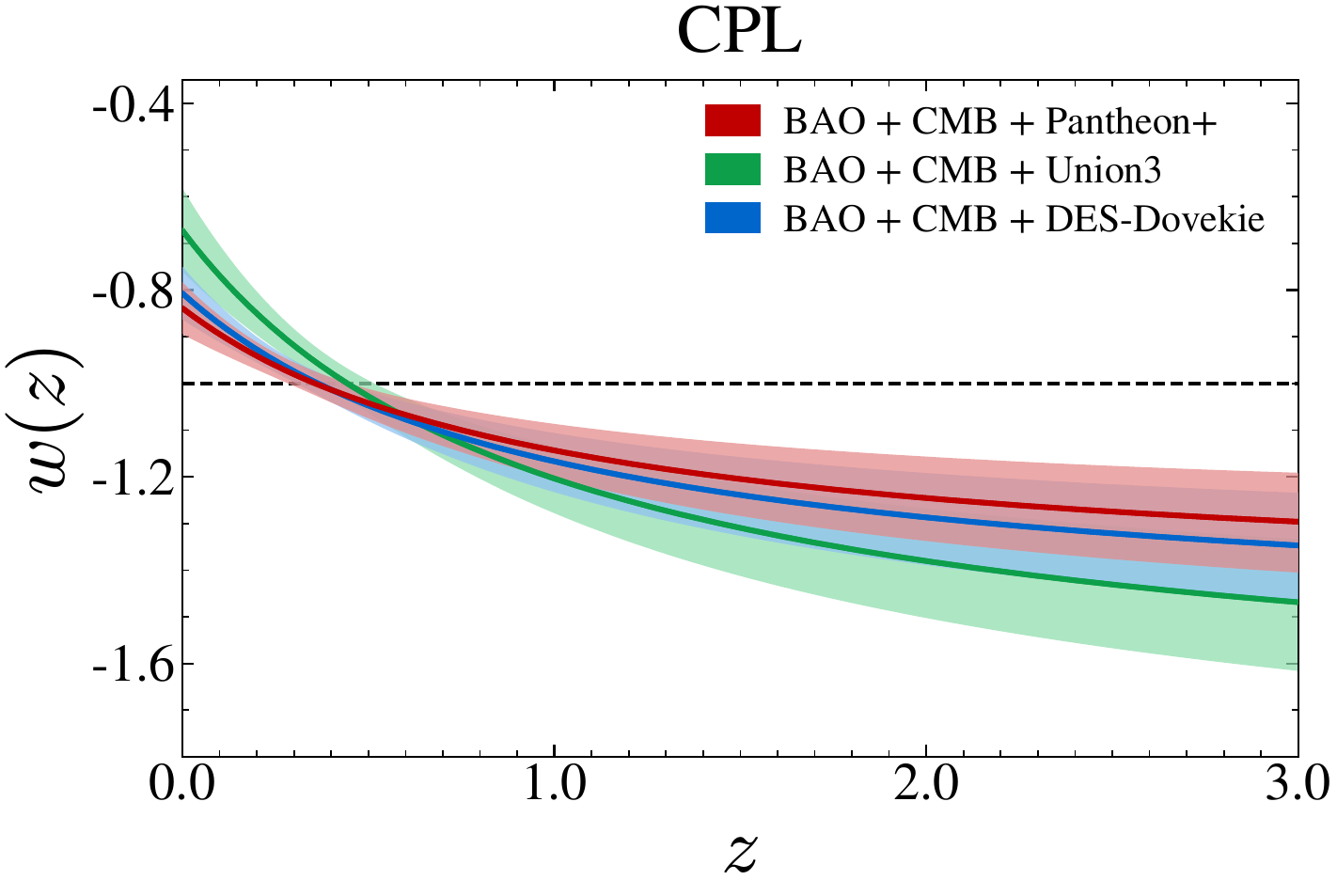}
\includegraphics[width=0.32\textwidth]{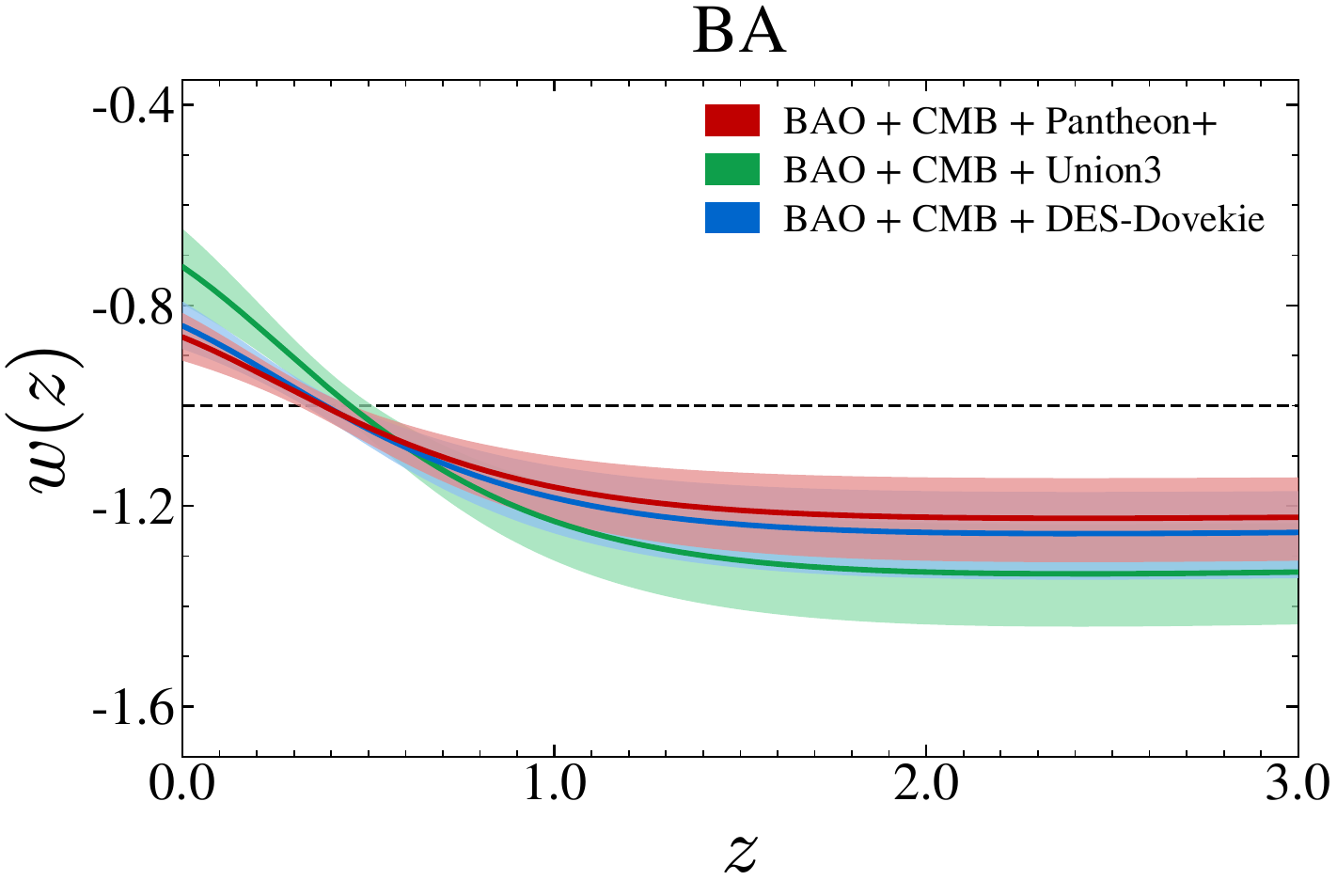}
\includegraphics[width=0.32\textwidth]{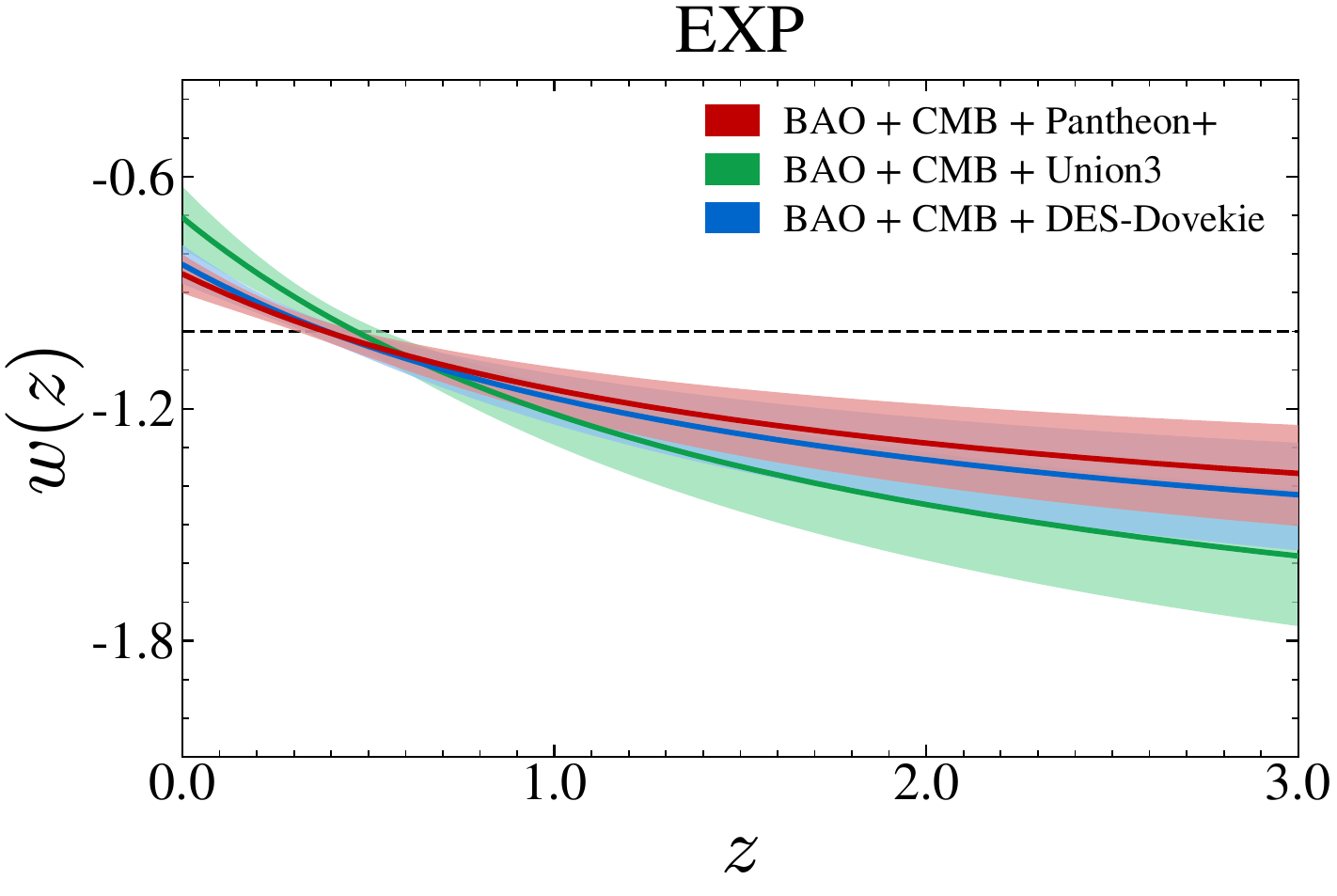}
\includegraphics[width=0.32\textwidth]{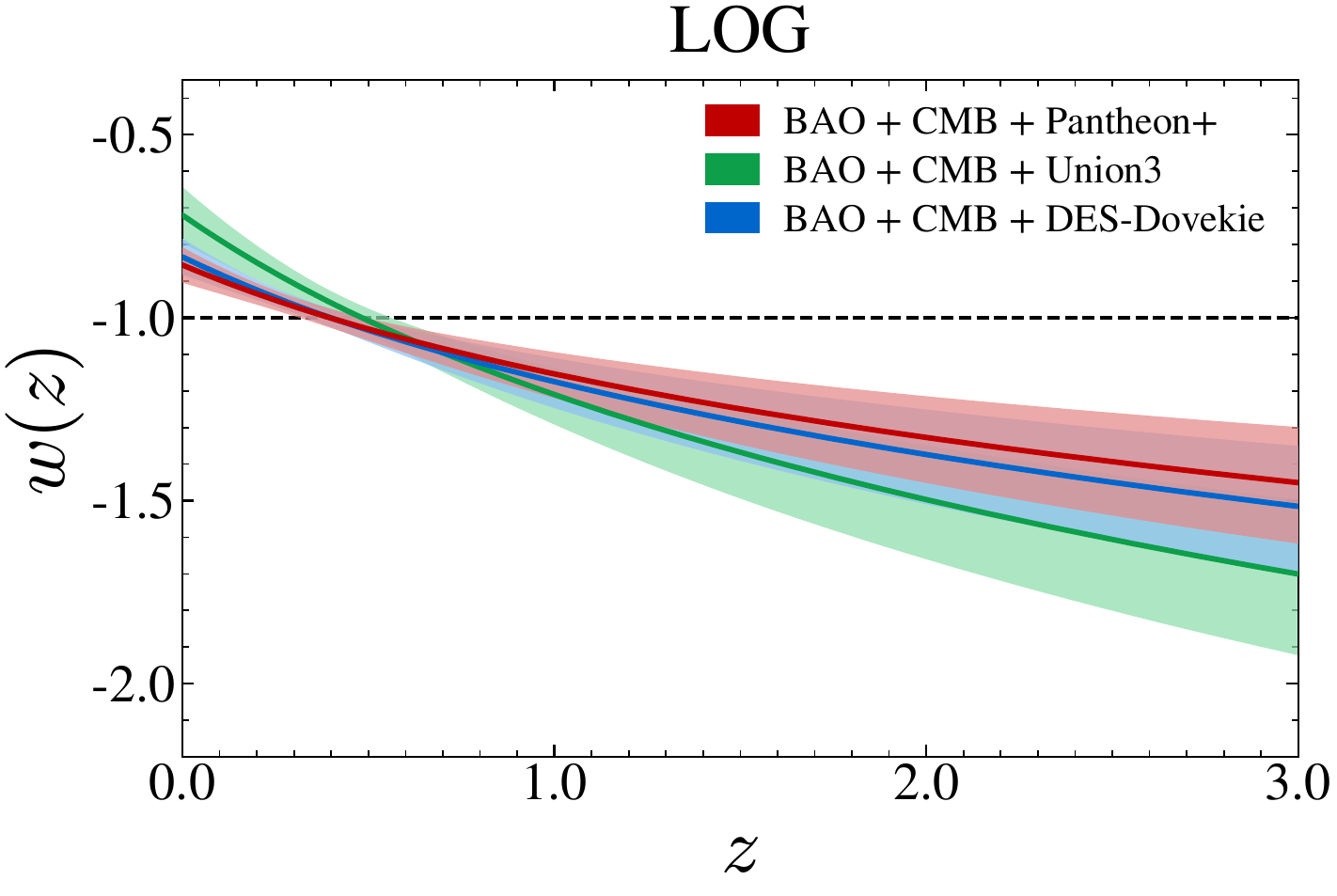}
\includegraphics[width=0.32\textwidth]{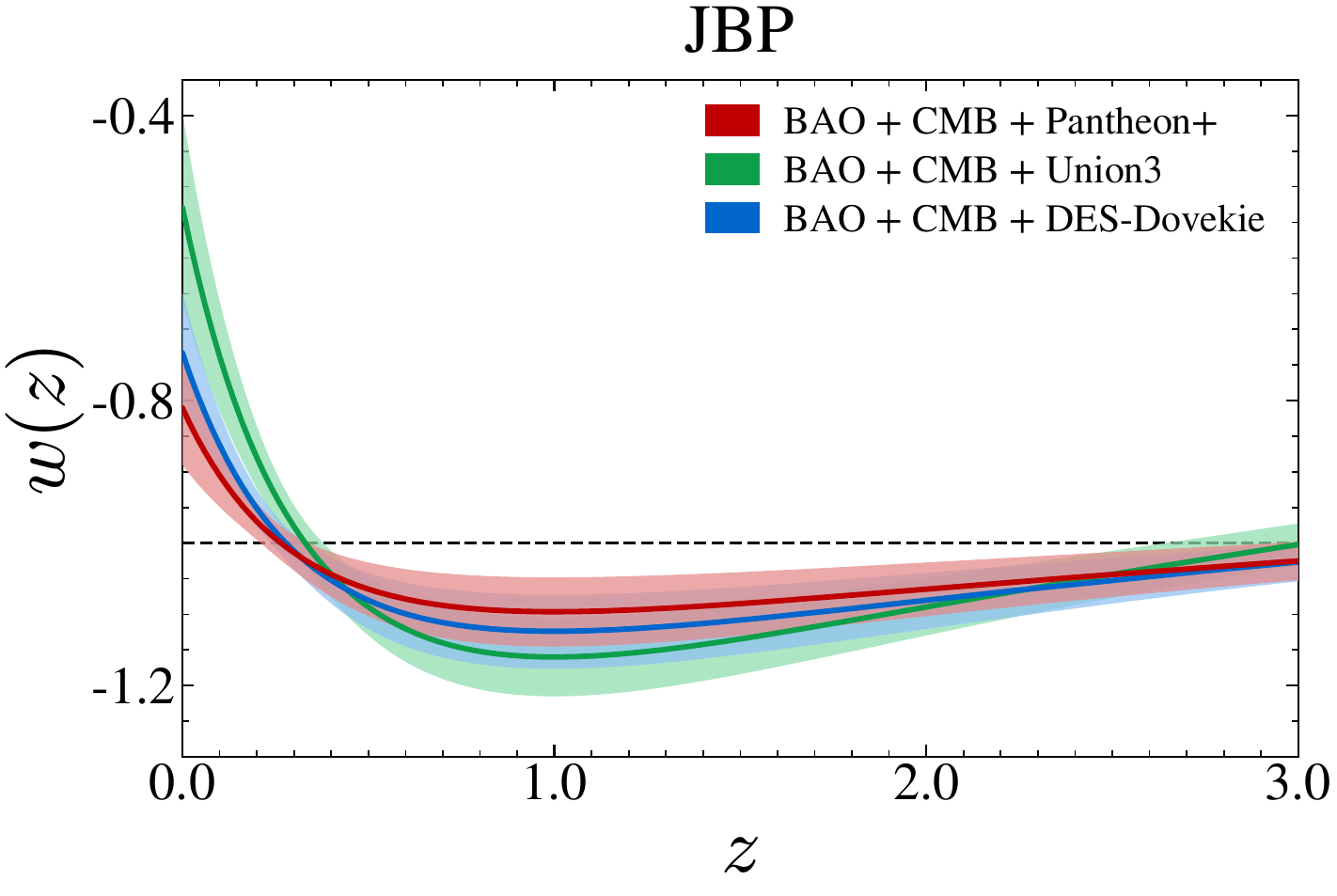}
\caption{
Evolution of the dark energy equation-of-state $w(z)$ for the CPL, BA, EXP, LOG, and JBP models. The red, green, and blue curves correspond to BAO+CMB+Pantheon+, BAO+CMB+Union3, and BAO+CMB+DES-Dovekie, respectively. The shaded regions denote the $68\%$ credible intervals, and the black dashed lines mark $w(z)=-1$.}
\label{fig:www}
\end{center}
\end{figure*}

\begin{figure*}[htbp]
\begin{center}
\includegraphics[width=0.32\textwidth]{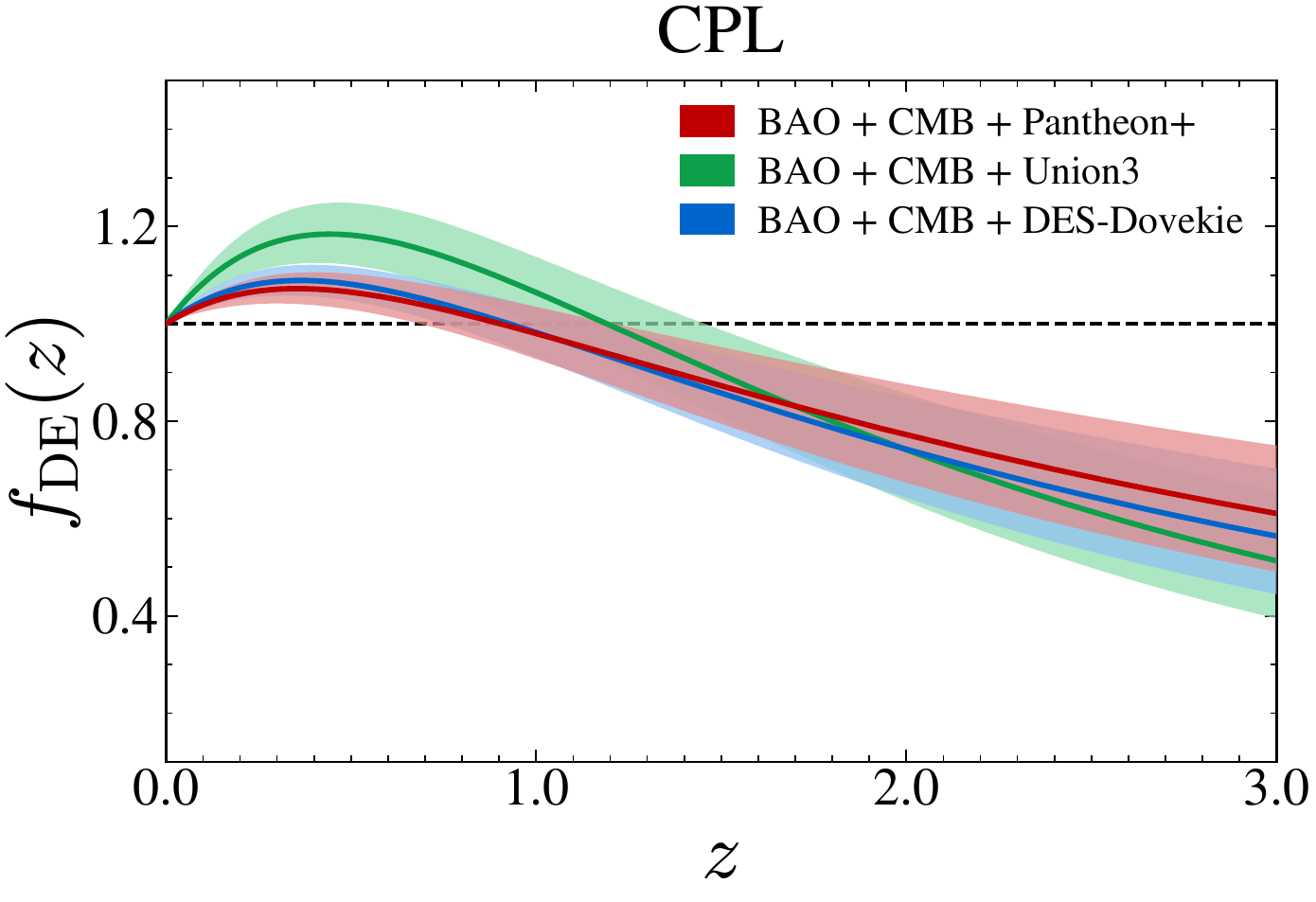}
\includegraphics[width=0.32\textwidth]{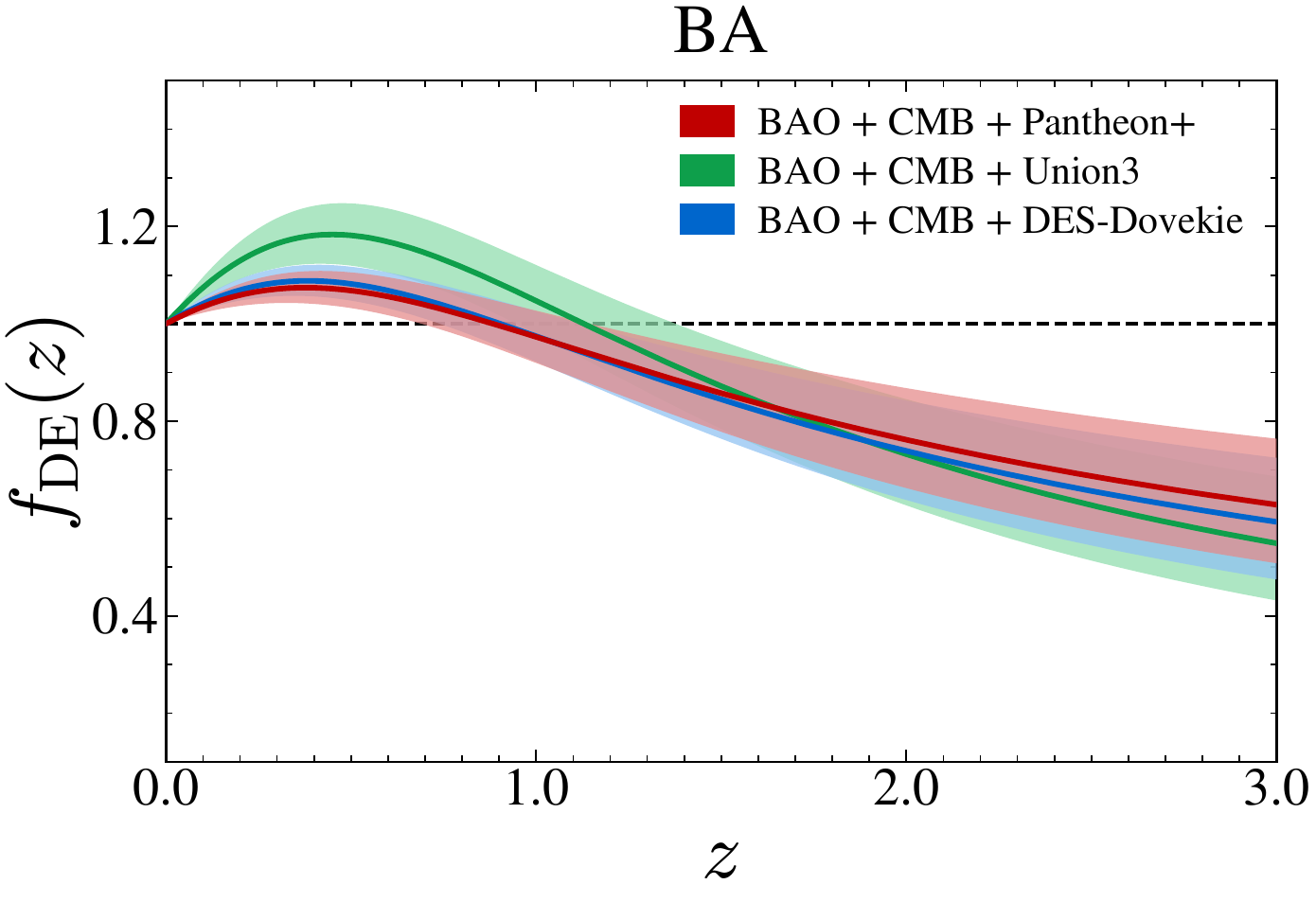}
\includegraphics[width=0.32\textwidth]{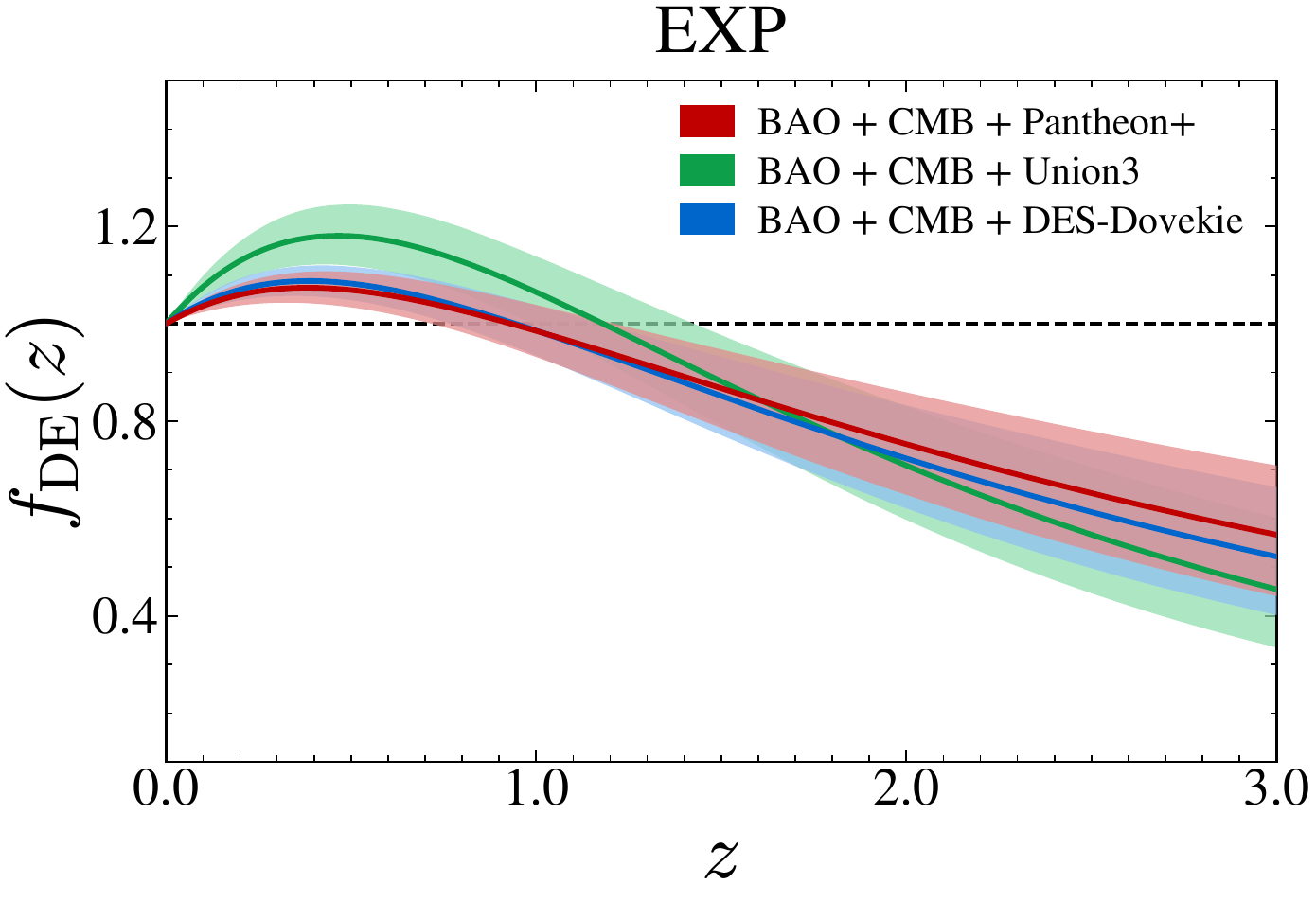}
\includegraphics[width=0.32\textwidth]{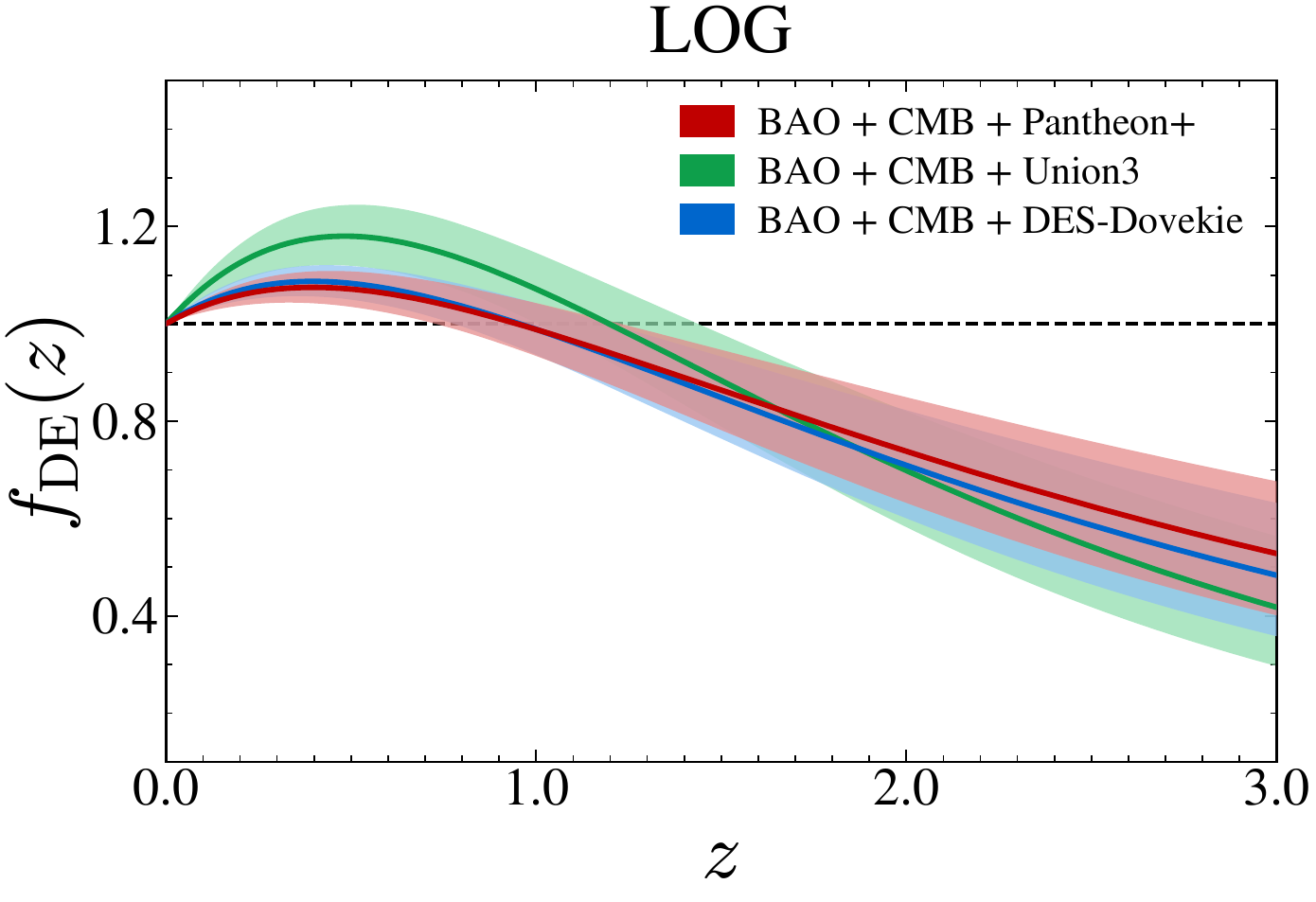}
\includegraphics[width=0.32\textwidth]{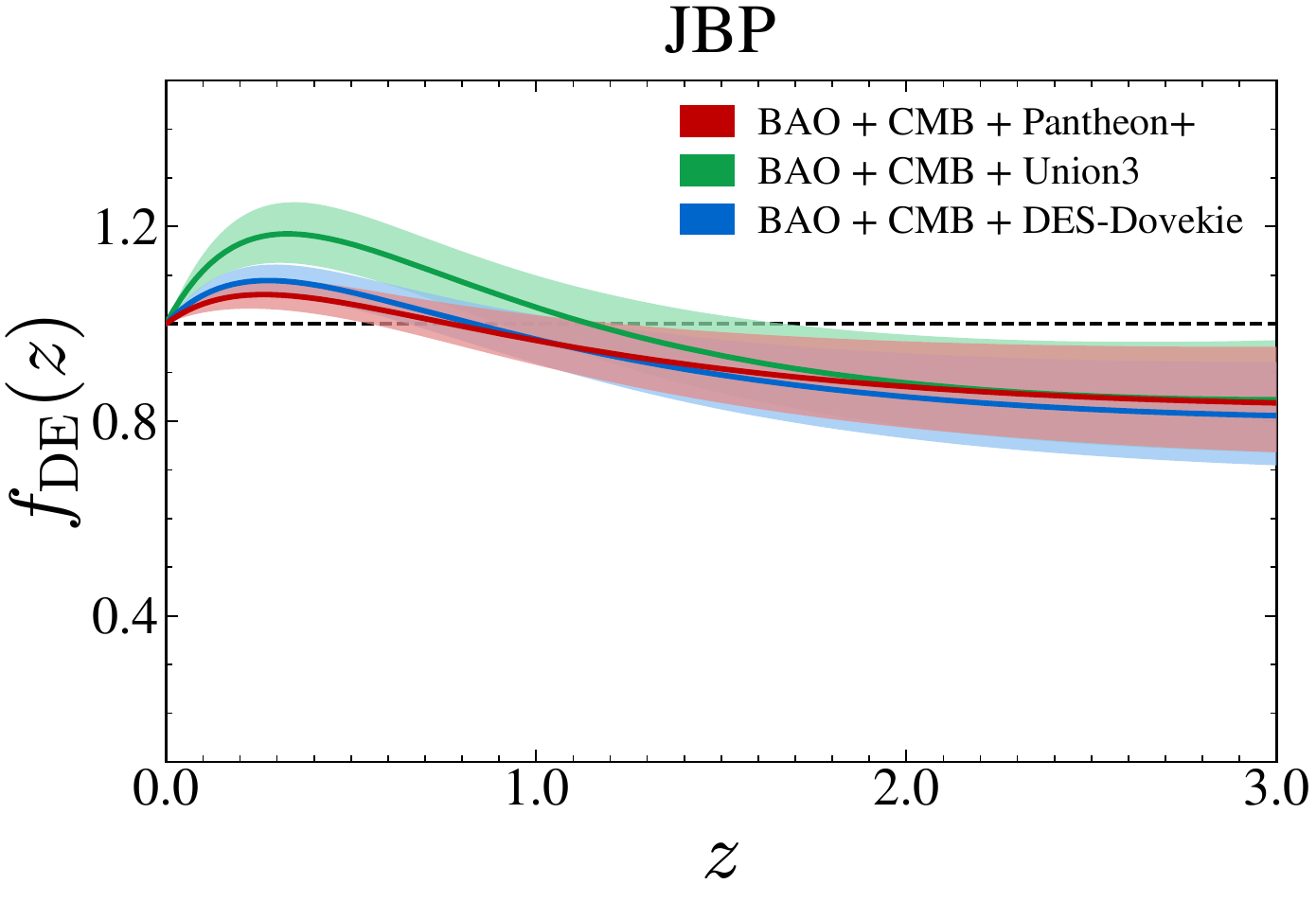}
\caption{Evolution of the normalized dark energy density $f_{\rm DE}(z)$ for the CPL, BA, EXP, LOG, and JBP models. The red, green, and blue curves correspond to BAO+CMB+Pantheon+, BAO+CMB+Union3, and BAO+CMB+DES-Dovekie, respectively. The shaded regions denote the $68\%$ credible intervals, while the black dashed lines represent the $\Lambda$CDM prediction $f_{\rm DE}(z)=1$.}
\label{fig:fff}
\end{center}
\end{figure*}

To show the evolution of dark energy more clearly, we reconstruct the evolution of $w(z)$ and $f_{\rm DE}(z)$ for the five parameterizations in Fig.~\ref{fig:www} and Fig.~\ref{fig:fff}. For the three SNe data combinations, all models satisfy
\begin{equation}
w(z=0)>-1.
\end{equation}
With the increase in redshift, $w(z)$ gradually decreases, crosses the $w(z)=-1$ boundary at $z\approx0.4$, then enters the phantom region. The crossing redshift is consistent with the position where $f_{\rm DE}(z)$ reaches its maximum in Fig.~\ref{fig:fff}, because $w(z)=-1$ corresponds to an extremum in the evolution of $f_{\rm DE}(z)$. This redshift dependent behavior can not be described by a constant equation-of-state, and shows that dark energy may evolve dynamically. The results for the JBP model shown in these figures also support an evolution from a quintessence-like state to a phantom-like state and then back to a quintessence-like state. Therefore, the difference is that the dark energy equation-of-state in the CPL, EXP, BA, and LOG models consistently exhibits a declining trend with the redshift increases, but the JBP model tends to stabilize at high redshifts. In all cases, $f_{\text{DE}}(z)$ are always equal to 1 at $z=0$, and the maximum values are also at $z\approx0.4$, which corresponds to the phantom crossing redshift. Although the specific evolutionary processes of these models vary, their overall trends in the evolution of dark energy are similar. These differences stem from the functional forms adopted by different parameterization models, but they do not alter the overall preference regarding the evolving $w(z)$ of dark energy. Similar results are obtained for the Pantheon+, Union3, and DES-Dovekie samples. This consistency shows that the evolution of dynamical dark energy does not depend on any specific SNe sample or specific parameterized model.

\begin{figure*}[htbp]
\begin{center}
\includegraphics[width=0.43\textwidth]{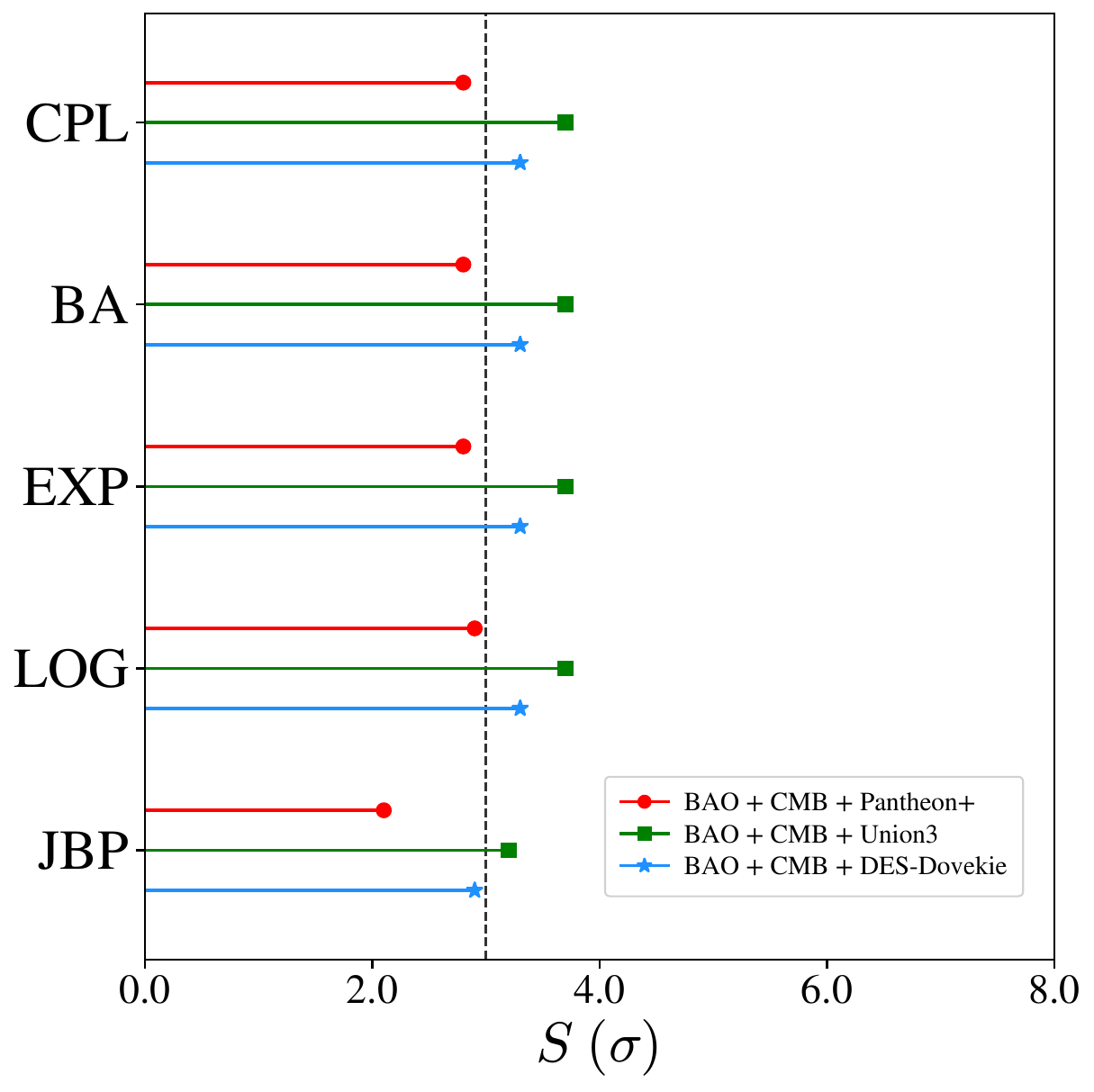}
\includegraphics[width=0.43\textwidth]{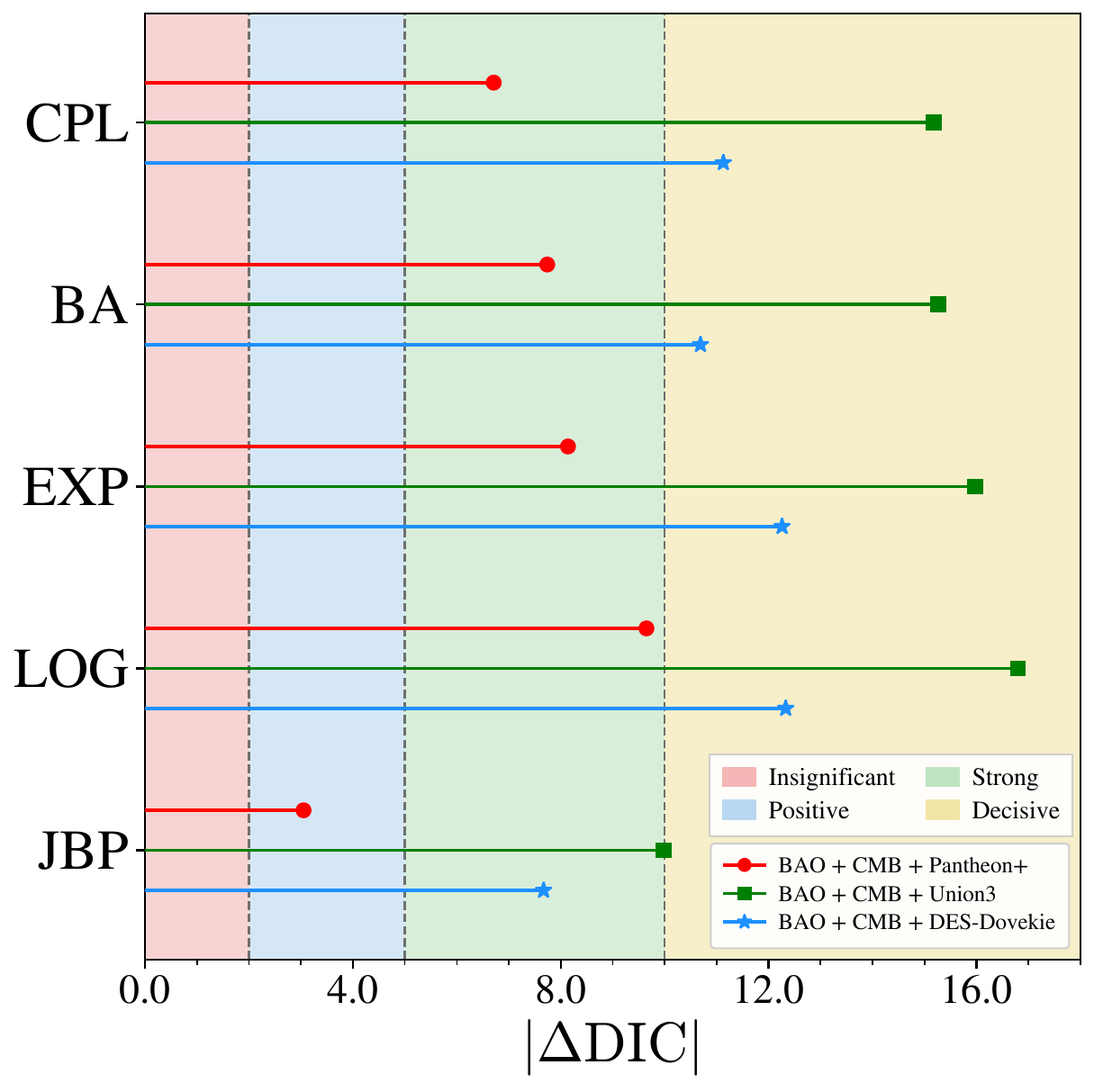}
\caption{The statistical significances $S$ (left) and $|\Delta\mathrm{DIC}|$ values (right) of the five dynamical dark energy models relative to the $\Lambda$CDM model for different combinations of BAO, CMB, and SNe data. The black dashed line in the left figure represents the position of $3\sigma$.}
\label{fig:S}
\end{center}
\end{figure*}

To evaluate the statistical preference for the five dynamical dark energy models relative to the $\Lambda$CDM model, we use two comparison methods: frequentist statistics~\cite{S} and DIC~\cite{DIC1, DIC2}. The results regarding the statistical significance of the improved fit and the difference in the DIC are shown in the Table~\ref{tab:S} and Fig.~\ref{fig:S}. We can see that the significances obtained by combining the Pantheon+ data give the lowest values, while the highest significances are obtained from the combined Union3 dataset. Furthermore, the significances obtained from the DES-Dovekie combination are all around $3\sigma$ for the five models, which are lower than that obtained using DES-SN5YR from DESI Collaboration~\cite{DESI2}, and consistent with the results from DES Collaboration~\cite{DES1, DES2}. Therefore, our results show that the current significances for the dynamical dark energy models are only around the evidence $3\sigma$, and can not reach $4\sigma$.

\begin{table}[htbp]
\renewcommand{\arraystretch}{1.2}
\begin{center}
\caption{Summary table of significance $S$ and $\Delta\mathrm{DIC}$ for various dynamical dark energy models relative to the $\Lambda$CDM.}
\label{tab:S}
\begin{tabular}{lccc}
  \hline
  \hline
  Model/Dataset & $S$ & $\Delta\mathrm{DIC}$ \\
  \hline
  \textbf{CPL} \\
  BAO + CMB + Pantheon+   & $2.8\sigma$ &$-6.7$ \\
  BAO + CMB + Union3      & $3.7\sigma$ & $-15.2$ \\
  BAO + CMB + DES-Dovekie \ \ \ \ & \ \ \ \ $3.3\sigma$ \ \ \ \ & \ \ \ \ $-11.1$ \ \ \ \ \\
  \textbf{BA} \\
  BAO + CMB + Pantheon+   & $2.8\sigma$ & $-7.7$ \\
  BAO + CMB + Union3      & $3.7\sigma$ & $-15.3$ \\
  BAO + CMB + DES-Dovekie & $3.3\sigma$ & $-10.7$ \\
  \textbf{EXP} \\
  BAO + CMB + Pantheon+   & $2.8\sigma$ & $-8.1$ \\
  BAO + CMB + Union3      & $3.7\sigma$ & $-16.0$ \\
  BAO + CMB + DES-Dovekie & $3.3\sigma$ & $-12.3$ \\
  \textbf{LOG} \\
  BAO + CMB + Pantheon+   & $2.9\sigma$ & $-9.7$ \\
  BAO + CMB + Union3      & $3.7\sigma$ & $-16.8$ \\
  BAO + CMB + DES-Dovekie & $3.3\sigma$ & $-12.3$ \\
  \textbf{JBP} \\
  BAO + CMB + Pantheon+   & $2.1\sigma$ & $-3.1$ \\
  BAO + CMB + Union3      & $3.2\sigma$ & $-10.0$ \\
  BAO + CMB + DES-Dovekie & $2.9\sigma$ & $-7.8$ \\
  \hline
  \hline
\end{tabular}
\end{center}
\end{table}

About the results of DIC, the statistical preferences for different models relative to $\Lambda$CDM based on the magnitude of $\left|\Delta\mathrm{DIC}\right|$ are categorized into four regions~\cite{DIC1, DIC2, DIC3}: the difference is ``insignificant" for $0\leq\left|\Delta\mathrm{DIC}\right|<2$, there is a ``positive" preference for $2\leq\left|\Delta\mathrm{DIC}\right|<5$, the preference is ``strong" for $5\leq\left|\Delta\mathrm{DIC}\right|<10$, and there is a ``decisive" preference for $\left|\Delta\mathrm{DIC}\right|\geq10$. We can see that, except for the JBP model with Pantheon+, whose result falls in the 0$\sim$5 range, the results for all other models are all above 5, indicating ``strong" preference for dynamical dark energy models. The main conclusions about the model preferences drawn from the $\left|\Delta\mathrm{DIC}\right|$ are similar to those drawn from the significances. The $\left|\Delta\mathrm{DIC}\right|$ values obtained from the Pantheon+ dataset combination are generally lower than those from the Union3 and DES-Dovekie combinations, whereas Union3 yields the highest $\left|\Delta\mathrm{DIC}\right|$ values across all five dynamical dark energy models. In brief, under the Union3 data combination, five models receive stronger evidence supporting dynamic dark energy, indicating that the choice of data combination indeed influences the model preferences. Overall, the results from all three datasets show a significant deviation from the $\Lambda$CDM model, indicating that dark energy likely evolves over time. But to draw definitive conclusion about the evolution of dark energy, more future data is desired.

\begin{figure*}[htbp]
\begin{center}
\includegraphics[width=0.47\textwidth]{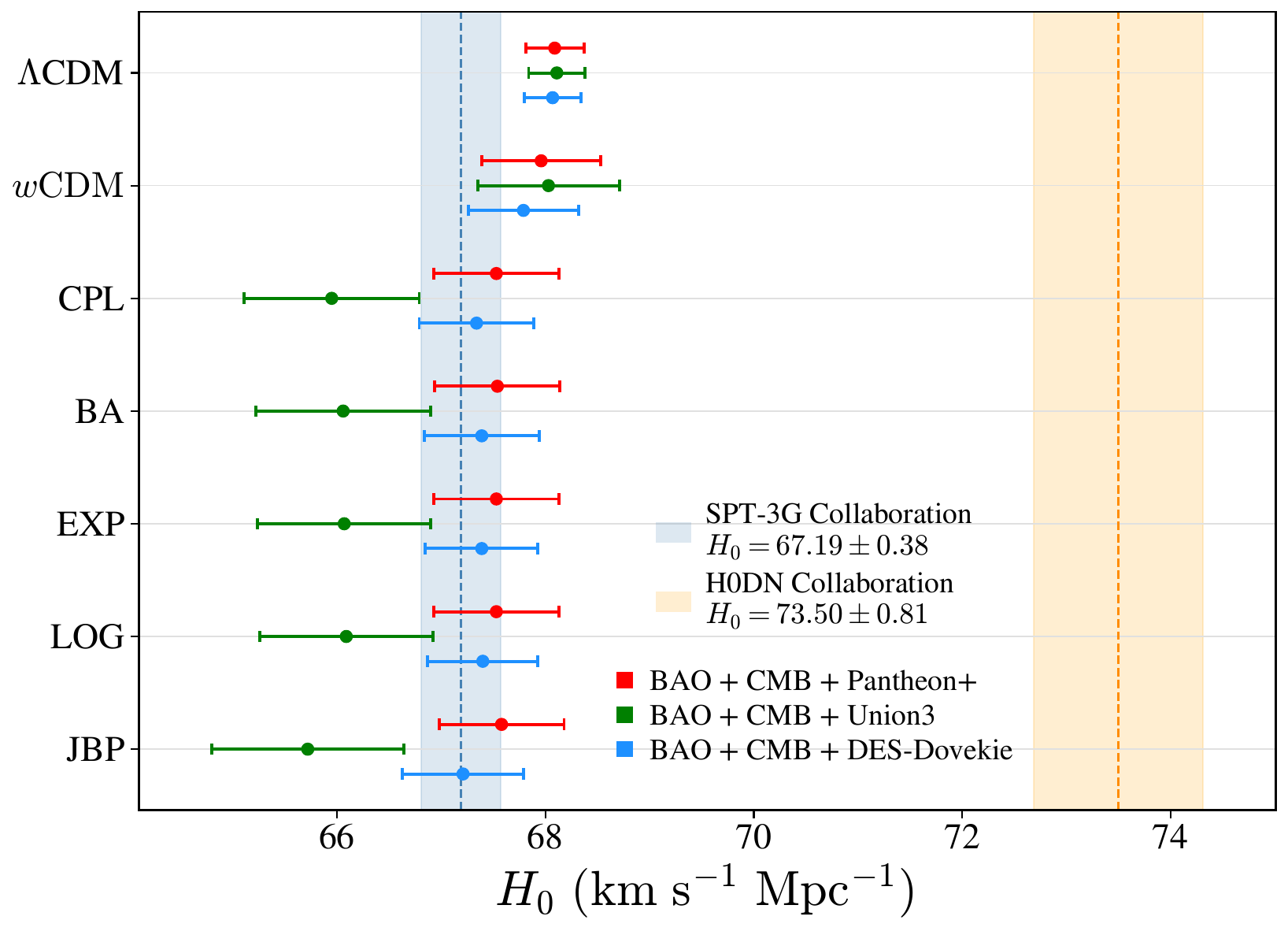}
\includegraphics[width=0.43\textwidth]{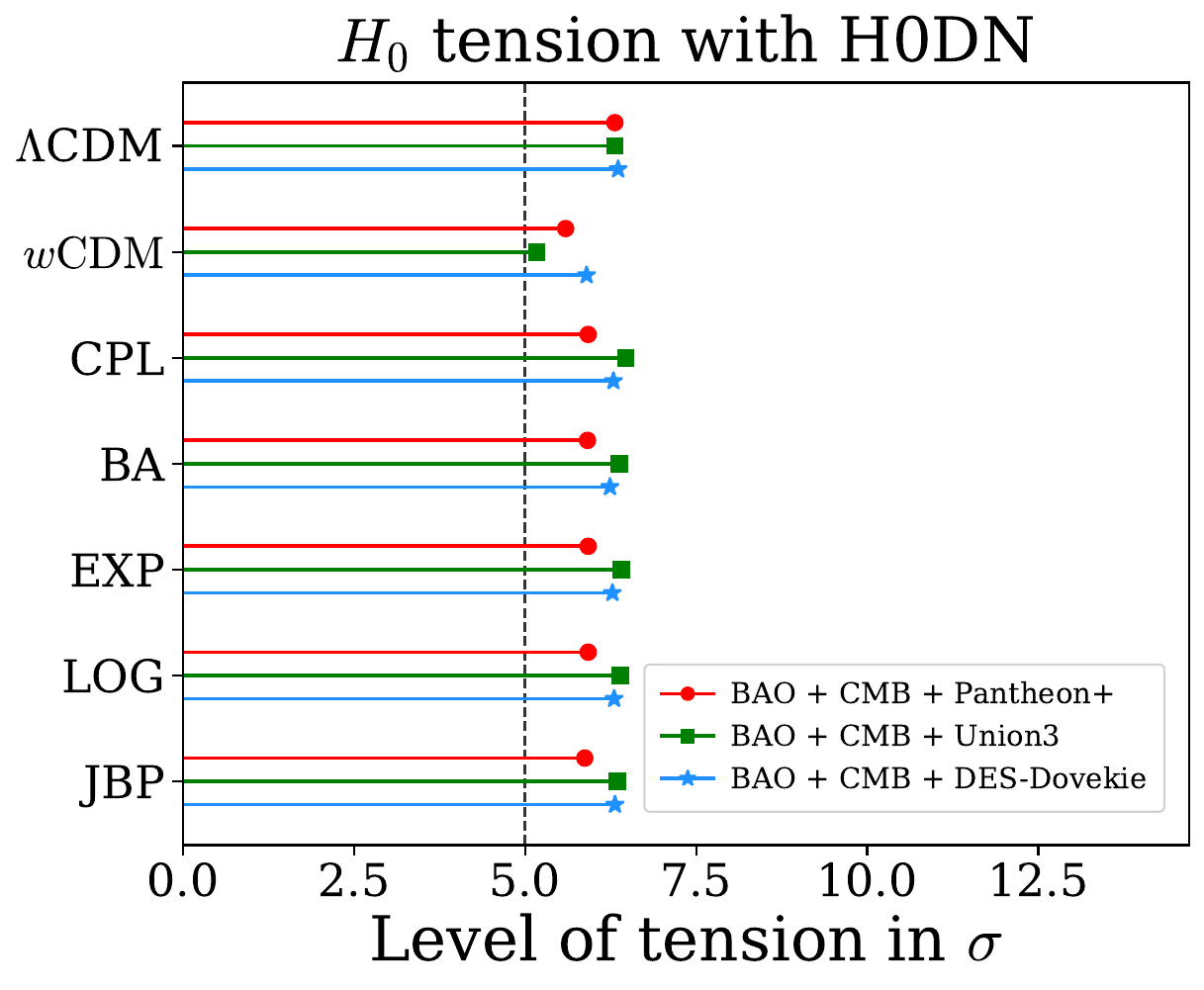}
\caption{Constraints on the Hubble parameter today $H_0$ in different dark energy models under various SNe data combinations (left), and the comparisons of the Hubble tension with H0DN measurement (right).}
\label{fig:H0}
\end{center}
\end{figure*}

The Hubble tension is one of the most widely discussed issues in modern cosmology~\cite{2019, 2022, 2021}. To investigate the impact of different dark energy models on this tension, we compare the $H_0$ constraints derived from the seven cosmological models in this wrok with direct measurements from the local universe and inferences regarding the early universe based on CMB observations. For the reference value of the local universe, we use the results obtained from the H0DN Collaboration~\cite{H03} using a distance network, $H_0=(73.50\pm0.81)$~km~s$^{-1}$~Mpc$^{-1}$. For the reference value of the early universe, we use the results obtained by the SPT-3G Collaboration~\cite{spt} through joint analysis of SPT, ACT, and \textit{Planck} CMB data, $H_0=(67.19\pm0.38)$~km~s$^{-1}$~Mpc$^{-1}$. The former represents direct measurement of the late-time universe based on distance-scale networks, whereas the latter is derived indirectly from CMB observations within the framework of $\Lambda$CDM cosmological model. By comparing the $H_0$ constraints derived from the seven models with these two reference values, we can test whether the dynamical dark energy models can alleviate the Hubble tension. The obtained $H_0$ constraints and the corresponding Hubble tension are shown in Fig.~\ref{fig:H0}. The measurement results in this work are all consistent with that of the SPT-3G Collaboration, whereas they differ significantly from the results of the H0DN Collaboration, where the tensions exceed $5\sigma$. Compared to the $\Lambda$CDM model, the dynamical dark energy models considered in this work do not significantly increase the central value of $H_0$, nor do they reduce the discrepancy with direct measurement from the local distance ladder. Therefore, based on the current results, these dynamical dark energy models can not effectively alleviate the Hubble tension.

\section{Conclusion}

In summary, we combine the latest DESI DR2 BAO data, complete CMB data, as well as Pantheon+, Union3, and DES-Dovekie SNe data to systematically compare the $\Lambda$CDM, $w$CDM, and five representative dynamic dark energy models. The results show that the CPL, EXP, BA, and LOG models generally exhibit quintom-B features evolving from past phantom-like state to current quintessence-like state, while the JBP model may undergo an evolution from quintessence-like to phantom-like and then back to quintessence-like state. The statistical significances for the dynamic dark energy models are $2.1\sim3.7\sigma$ for different SNe samples and different parametrization models. Among them, Union3 gives the highest significances for the deviation from $\Lambda$CDM, while Pantheon+ shows a relatively weaker preference, and the JBP model gives the lowest significances compared to the other dynamic dark energy models. In addition, the $H_0$ values obtained from each model are still closer to the CMB inference results from the early universe, and there is a clear difference from the direct measurements in the local universe, indicating that the dynamic dark energy models considered here can not effectively alleviate the Hubble tension. Overall, the current measurements generally favour dynamic dark energy at the level of around $3\sigma$. But because the significance is only around the evidence $3\sigma$, it is still uncertain about the evolution of dark energy. Therefore, in the future, with the release of new measurement data and the advent of more precise measurement methods, the uncertainties of cosmological parameters can be further reduced. This will allow us to check whether the dynamic dark energy really exists and deepen the understanding of its properties, ultimately revealing the physical nature of dark energy.

\section{ACKNOWLEDGMENTS}
This work is supported by the Natural Science Foundation of Henan under Contract No. 242300421163; and the National Natural Science Foundation of China (NSFC) under Contracts Nos. 12375071.

\end{document}